\documentclass[preprint,norevision]{ptephy_om}
\preprintnumber{RIKEN-iTHEMS-Report-26}
\revisionnum{1}

\usepackage{hyperref}
\usepackage{orcidlink}
\usepackage{booktabs}
\usepackage{longtable}
\usepackage{array}

\DeclareMathOperator{\Det}{Det}
\DeclareMathOperator{\Ln}{Ln}

\title{Center-twisted Gribov spectra and the finite-volume Gaussian response
in the refined Gribov--Zwanziger framework}

\author{Okuto Morikawa \orcidlink{0000-0002-0044-4491}}
\affil{Center for Interdisciplinary Theoretical and Mathematical Sciences (iTHEMS),
RIKEN, Wako 351-0198, Japan
\email{okuto.morikawa@riken.jp}}

\begin{document}

\begin{abstract}
We develop a continuum framework for comparing the Gribov--Zwanziger and center-vortex descriptions of confinement through the gauge-invariant twisted partition function of the electric $\mathbb{Z}_N^{[1]}$ 1-form symmetry. A background 2-form field $B$, equivalently an 't~Hooft twist on a torus, labels a global sector and is not itself a dynamical center vortex.

For a minimal irreducible twist on $T^4$, we derive the complete adjoint momentum lattice of $SU(N)$. The twisted spectrum is exactly the scalar spectrum on an enlarged torus with periods $(L_1,L_2,NL_3,NL_4)$ with the ordinary-torus sublattice removed. This yields a finite Faddeev--Popov gap at the flat representative and reduces twisted-minus-untwisted spectral traces to ordinary torus traces. For the refined Gribov--Zwanziger (RGZ) kernel, Poisson resummation gives an exact finite-volume Bessel-function winding sum with a universal center-twist projector.

We evaluate the Gaussian one-loop integral at fixed RGZ parameters up to the finite-dimensional global zero-mode/stabilizer normalization. The Zwanziger determinants cancel, while the gauge-fixing/ghost sector leaves a universal massless primed determinant on $T^4$; the untwisted sector also contains constant gluon modes. The only normalization not fixed by the local quadratic Hessian is the relative zero-mode/stabilizer measure of the reducible untwisted and irreducibly twisted flat connections. We also derive closed finite-volume sources for the RGZ stationary equations.

The massive response is exponentially suppressed at large volume, whereas the massless factor depends on the global zero-mode normalization. Thus the Gaussian calculation does not by itself establish the strong center-vortex-condensation criterion for $Z[B]/Z[0]$, but it isolates the remaining global normalization problem and separates a global twist from an actual dynamical vortex.
\end{abstract}

\maketitle

\tableofcontents

\section{Introduction}
\label{sec:intro}

Color confinement in Yang--Mills theory admits several complementary
nonperturbative descriptions.  Two particularly persistent pictures start
from rather different structures.  The Gribov--Zwanziger (GZ) framework
addresses the failure of the Faddeev--Popov construction to fix non-Abelian
gauge fields globally.  In Landau gauge, the functional integral is restricted
to the first Gribov region, where the Faddeev--Popov operator is positive
\cite{Gribov:1977wm,Zwanziger:1989mf,Vandersickel:2012tz}.  The global
obstruction to choosing one representative continuously on every gauge orbit
was sharpened by Singer \cite{Singer:1978dk}.  The refined
Gribov--Zwanziger (RGZ) framework incorporates dimension-two condensates and
leads to an infrared gluon propagator of decoupling type, in quantitative
agreement with lattice-motivated infrared behavior \cite{Dudal:2008sp}.

The center-vortex picture instead emphasizes extended magnetic defects.
Their linking with Wilson loops produces center phases and naturally
organizes the $N$-ality dependence expected in a confining phase
\cite{Greensite:2003bk}.  A relation between center vortices and the Gribov
horizon has long been suggested by gauge-fixed studies.  On the lattice,
removing center vortices removes the characteristic enhancement of low-lying
Faddeev--Popov modes, and thin center vortices occupy singular loci of the
Gribov horizon \cite{Greensite:2004ur}.  In the continuum, additional
Faddeev--Popov zero modes have been found in center-vortex backgrounds
\cite{Maas:2005ym}.  These results are striking, but the comparison is not yet
organized around a gauge-invariant observable.

Generalized global symmetry provides such an observable.  Pure
$SU(N)$ Yang--Mills theory has an electric $\mathbb{Z}_N^{[1]}$
1-form symmetry, which can be coupled to a background 2-form gauge field $B$
\cite{Gaiotto:2014kfa}.  On a torus, a nontrivial class of $B$ is equivalently
represented by an 't~Hooft twist \cite{tHooft:1979rtg}.  The corresponding
partition function $Z[B]$ is gauge invariant.  Hayashi and Tanizaki recently
proposed the torus twisted partition function as a gauge-invariant criterion
for center-vortex condensation \cite{Hayashi:2026ht}.  For the minimal
nontrivial twist, their weak criterion is
\begin{equation}
 \frac{Z_{T^4}[B]}{Z_{T^4}[0]}
 \longrightarrow C_B,
 \qquad
 C_B\neq0,
 \label{eq:weakcriterion}
\end{equation}
in the infinite-volume limit, whereas the strong criterion requires
\begin{equation}
 \frac{Z_{T^4}[B]}{Z_{T^4}[0]}
 \longrightarrow 1.
 \label{eq:strongcriterion}
\end{equation}
This makes it possible to ask about vortex condensation without defining an
individual vortex by maximal-center gauge or another projection procedure.

The central point of the present work is that three logically different
objects should not be conflated:
\begin{enumerate}
 \item the background field $B$, which labels a 1-form-symmetry-twisted
 sector;
 \item the Gribov region in that sector, defined by the spectrum of the
 sectorwise Faddeev--Popov operator;
 \item a dynamical center-vortex configuration contributing to the path
 integral.
\end{enumerate}
The first is gauge invariant but nondynamical, the second is a property of a
gauge-fixed functional measure, and the third is a field configuration.
Their relation is a dynamical statement and should be tested rather than
built into the definitions.

The ingredients entering this problem have a long history, but they developed
along largely distinct lines.  In particular, the color-momentum basis,
effective enlarged periods, and perturbative finite-volume analysis of
irreducibly twisted tori are established ingredients of twisted-box
perturbation theory and volume reduction
\cite{GarciaPerez:2014vt,Bribian:2019wqj}.  We use these results as input and
do not claim the shifted adjoint lattice, the enlarged-torus interpretation,
or Poisson resummation separately as new.  Our narrower contribution is to
combine them with a sectorwise Gribov/RGZ measure whose primary target is the
one-form-symmetry-twisted ratio $Z[B]/Z[0]$, while keeping the flat bundle and
a dynamical vortex saddle conceptually distinct.  Appendix~\ref{app:history}
states this novelty boundary explicitly and gives a historical map of the
ingredients.

We take a first step in this direction by studying the Gribov/RGZ spectral
problem on $T^4$ with a fixed irreducible center twist.  The flat local
representative is particularly useful: all topology is carried by transition
functions, so every adjoint field is expanded on an explicitly shifted
momentum lattice.  This observation gives two immediate results.  First, an
irreducible twist removes the continuous global-color zero modes and generates
a finite spectral gap of the free Faddeev--Popov operator.  Thus the existence
of nontrivial 't~Hooft flux does \textit{not} by itself place the configuration
on the Gribov horizon.  Second, the finite-volume response of any quadratic
kernel can be evaluated by standard heat-kernel methods.

We carry this out for the transverse RGZ kernel.  The resulting
center-twisted minus untwisted determinant can be Poisson-resummed in closed
form.  For a minimal irreducible $SU(N)$ twist, the entire color
structure reduces to a simple arithmetic condition on winding numbers.  This
provides an analytic finite-volume observable that can be evaluated before a
localized vortex profile is introduced.  In the RGZ regime with nonzero
screening scales, the response is exponentially suppressed at large volume.
When the RGZ poles occur as a complex-conjugate pair, the corresponding terms
can show damped oscillations as functions of the box size.

The transverse result can in fact be completed considerably further at the
Gaussian level.  Imposing Landau gauge exactly, the delta functional removes
the longitudinal gluon amplitude with a Jacobian $z^{-1/2}$ for each nonzero
scalar eigenvalue $z$, while the Faddeev--Popov determinant contributes $z$.
They therefore leave a residual factor
$(\Det'[-\partial_B^2])^{1/2}$.  The bosonic and Grassmann Zwanziger
localizing fields have identical operator and multiplicity, so their
normalization determinants cancel; the $A$--$\varphi$ mixing generates the
RGZ transverse kernel.  This is the finite-volume counterpart of the
standard one-loop RGZ construction, in which the remaining massless
determinant is scaleless on $\mathbb{R}^d$ and is normally absent from the
reported vacuum energy \cite{Dudal:2008sp}.

On a torus the massless factor cannot be dropped.  We evaluate it as a primed
zeta determinant and show that the complete nonzero-mode Gaussian response is
known exactly once the transverse result is known.  The twisted sector has no
adjoint zero mode, whereas the untwisted flat representative has both
constant gluon modes and continuous global-color gauge zero modes.  The
constant gluons can also be included explicitly because the RGZ kernel is
finite at $p=0$.  What remains outside the local quadratic Hessian is the
normalization of the different global stabilizers.  This is a genuine
reducible-versus-irreducible flat-connection issue rather than another
functional determinant.  A common microscopic measure, or an explicitly
fixed global gauge prescription, is needed to assign this last factor.

The localization must moreover be globally consistent on the twisted bundle.
Background-gauge-invariant GZ constructions make clear that the auxiliary
fields cannot simply be assigned ad hoc periodic boundary conditions; their
transformation law has to reproduce the globally defined horizon functional
\cite{Kroff:2018,Capri:2016aqq,Justo:2022eyg}.  Our determinant cancellation
should therefore be understood as a statement about any such consistent
localization of the sectorwise nonlocal kernel.

The second stage is local rather than global.  On
$\mathbb{R}^2\times T^2$ with 't~Hooft flux, semiclassical center vortices can
be studied in a weakly coupled regime \cite{Tanizaki:2022ngt}.  Evaluating the
Faddeev--Popov spectrum around such a saddle in a fixed twisted sector gives a
direct modern version of the vortex--Gribov-horizon problem.  The combination
of the torus partition-function response and the vortex-saddle spectrum is the
bridge we advocate.

The paper is organized as follows.  Section~\ref{sec:center} reviews the
background 2-form field and the twisted partition-function observable.
Section~\ref{sec:gribov} defines the Gribov problem sector by sector.
Section~\ref{sec:twist} derives the irreducible $SU(N)$ adjoint
momentum lattice and its Faddeev--Popov gap.  Section~\ref{sec:rgz} contains
the main calculation: the exact Poisson-resummed transverse RGZ response.
Section~\ref{sec:completion} completes the Gaussian functional integral,
including the massless gauge determinant, constant gluon modes, and the
sectorwise gap equations, and isolates the remaining global-stabilizer
normalization.
Section~\ref{sec:vortex} connects the construction to an actual vortex saddle.
Section~\ref{sec:numerics} presents numerical checks and a practical route to a
quantitative test, and Sec.~\ref{sec:conclusion} summarizes the implications.
Appendix~\ref{app:poisson} gives the Poisson-resummation derivation of the shifted determinant, Appendix~\ref{app:color} derives the color character of the irreducible twist, Appendix~\ref{app:gaussian} collects the Gaussian determinant counting, Appendix~\ref{app:massless} derives the scale dependence of the massless primed determinant, and Appendix~\ref{app:history} states the scope of the novelty claim and summarizes the historical development.

\section{Center symmetry and twisted partition functions}
\label{sec:center}

\subsection{Background 2-form field}

Let $M_4$ be a closed Euclidean four-manifold.  The
$\mathbb{Z}_N^{[1]}$ center symmetry of pure $SU(N)$ Yang--Mills
theory can be coupled to a background 2-form gauge field $B$.  A useful
continuum description promotes the gauge connection to a $U(N)$ connection
$\widetilde a$ and imposes
\begin{equation}
 N B = \tr \widetilde F,
 \label{eq:NB}
\end{equation}
where $\widetilde F$ is the $U(N)$ field strength.  Under the 1-form gauge
redundancy,
\begin{align}
 \widetilde a &\longrightarrow
 \widetilde a+\lambda\bm{1}_N,
 \\
 B &\longrightarrow B+\rmd\lambda,
 \label{eq:oneformgauge}
\end{align}
so that the traceless combination of the field strength is invariant.  The
important point for the present paper is that $B$ is nondynamical.  It is a
probe specifying how the theory is placed in a nontrivial 1-form-symmetry
background.

On a rectangular four-torus with periods $L_\mu$, the same information can be
encoded by 't~Hooft fluxes
\begin{equation}
 n_{\mu\nu}\in\mathbb{Z}_N,
 \qquad
 \frac{1}{2\pi}
 \int_{(T^2)_{\mu\nu}} B
 =\frac{n_{\mu\nu}}{N}
 \quad (\mathrm{mod}\ 1).
 \label{eq:Bflux}
\end{equation}
Gauge fields are then specified by transition functions $\Omega_\mu$ obeying
the projective cocycle condition
\begin{equation}
 \Omega_\mu(x+L_\nu\hat\nu)\Omega_\nu(x)
 =z_{\mu\nu}\,
 \Omega_\nu(x+L_\mu\hat\mu)\Omega_\mu(x),
 \label{eq:twistcocycle}
\end{equation}
with
\begin{equation}
 z_{\mu\nu}
 =\exp\!\left(\frac{2\pi\rmi n_{\mu\nu}}{N}\right).
 \label{eq:zmunu}
\end{equation}
Adjoint fields do not see the center element directly, but they do see the
adjoint action of the transition functions.  This is the origin of the
fractionally shifted momenta derived below.

\subsection{The primary gauge-invariant observable}

For a fixed background $B$, define
\begin{equation}
 R[B]
 =\frac{Z_{M_4}[B]}{Z_{M_4}[0]},
 \label{eq:Rdef}
\end{equation}
and
\begin{equation}
 \Delta F[B]
 =-\ln R[B].
 \label{eq:DeltaF}
\end{equation}
For the minimal torus twist, Eqs.~\eqref{eq:weakcriterion} and
\eqref{eq:strongcriterion} provide the weak and strong center-vortex
condensation criteria of Ref.~\cite{Hayashi:2026ht}.

The Gribov/RGZ construction must ultimately provide a representation of this
same gauge-invariant ratio.  It is therefore useful to distinguish the exact
quantity $R[B]$ from intermediate gauge-fixed quantities.  We shall denote a
contribution obtained from a specific quadratic RGZ sector by a subscript, for
example $R_T[B]$ for the transverse gluon kernel.  Agreement of an
intermediate quantity with Eq.~\eqref{eq:strongcriterion} is a consistency
property, not by itself a proof of the physical criterion.

\section{The Gribov problem in a fixed twisted sector}
\label{sec:gribov}

\subsection{Sectorwise Gribov region}

Let $\mathcal{A}_B$ be the space of connections in the bundle selected by $B$
and let $\mathcal{G}_B$ be the corresponding group of admissible gauge
transformations.  In a local trivialization, Landau gauge has its familiar
form,
\begin{equation}
 \partial_\mu A_\mu=0,
 \label{eq:Landau}
\end{equation}
but the fields and gauge parameters satisfy the twisted patching conditions.
The Faddeev--Popov operator is
\begin{equation}
 \mathcal{M}_B^{ab}[A]
 =-\partial_\mu D_\mu^{ab}[A],
 \label{eq:FPB}
\end{equation}
acting on sections of the adjoint bundle.

We define the first Gribov region sector by sector,
\begin{equation}
 \Omega_B
 =\left\{
 A\in\mathcal{A}_B\,\middle|\,
 \partial_\mu A_\mu=0,
 \ \mathcal{M}_B[A]>0
 \right\},
 \label{eq:OmegaB}
\end{equation}
where any genuine gauge zero modes are removed in defining the positivity
condition.  Formally, the restricted partition function is
\begin{equation}
 Z_\Omega[B]
 =\int_{\Omega_B}\mathcal{D}A\,
 \delta(\partial_\mu A_\mu)
 \Det'\mathcal{M}_B[A]\,
 \exp\{-S_{\mathrm{YM}}[A]\}.
 \label{eq:ZOmega}
\end{equation}
Equation~\eqref{eq:ZOmega} is a useful definition even before the horizon
functional is localized.  In particular, it makes clear that the domain of
the Faddeev--Popov operator, and not only its local differential expression,
depends on $B$.

\subsection{Why a nontrivial twist need not lie on the horizon}
\label{subsec:flatgap}

For the irreducible twists used below, the topology can be placed entirely in
constant transition functions and the local connection can be chosen as
\begin{equation}
 A_\mu^{(0)}=0.
 \label{eq:flatA}
\end{equation}
The Faddeev--Popov operator at this representative is simply
\begin{equation}
 \mathcal{M}_B[A^{(0)}]=-\partial^2,
 \label{eq:FPflat}
\end{equation}
but its domain is twisted.  As we will show explicitly, every adjoint color
component is nonperiodic in at least one of the twisted directions.  Hence
there is no nonzero constant adjoint gauge parameter compatible with both
transition functions.

This observation is conceptually important.  A center-twisted sector is not
synonymous with a configuration on the Gribov horizon.  At the flat
representative the twist can instead move the free Faddeev--Popov spectrum
\textit{away} from zero.  Horizon contact, if generated by a dynamical center
vortex, is therefore additional information about a particular field
configuration in the sector.

\revone{Here, the flat representative refers specifically to
$A_\mu^{(0)}=0$ with the chosen constant transition functions.
The global bundle data encoded by the twist are retained exactly
through the boundary conditions, while the fluctuation expansion
is performed about this particular representative.
This expansion is local in configuration space, not a
short-distance approximation on the spacetime torus:
the quadratic spectra retain their full finite-volume dependence
on the specified twist.
This should be distinguished from an analysis of the entire
gauge orbit.
Since the Landau-gauge Faddeev--Popov operator is not gauge
covariant, its positivity at this representative does not imply
positivity at every gauge-equivalent representative satisfying
Landau gauge.}

\section{Irreducible minimal twist and adjoint momentum lattice}
\label{sec:twist}

\subsection{Twist eaters for \texorpdfstring{$SU(N)$}{SU(N)}}

We place a minimal flux through the $(3,4)$ torus,
\begin{equation}
 n_{34}=1\quad(\mathrm{mod}\ N),
 \qquad
 n_{\mu\nu}=0
 \quad\text{otherwise}.
 \label{eq:minimalflux}
\end{equation}
Let
\begin{equation}
 \omega=\exp\!\left(\frac{2\pi\rmi}{N}\right).
 \label{eq:omega}
\end{equation}
Up to irrelevant center phases needed to choose representatives in
$SU(N)$, one may use the clock and shift matrices $P$ and $Q$ as
constant transition functions,
\begin{equation}
 \Omega_3=P,
 \qquad
 \Omega_4=Q,
 \qquad
 P Q=\omega Q P.
 \label{eq:PQ}
\end{equation}
A convenient basis of the complexified adjoint algebra is
\begin{equation}
 J_{r,s}
 \propto P^r Q^s,
 \qquad
 (r,s)\in\mathbb{Z}_N^2,
 \qquad
 (r,s)\neq(0,0).
 \label{eq:Jrs}
\end{equation}
The omitted $(0,0)$ element is proportional to the identity and is not part of
$\mathfrak{su}(N)$.

The adjoint actions are diagonal in this basis,
\begin{align}
 \operatorname{Ad}(P)J_{r,s}
 &=\omega^s J_{r,s},
 \\
 \operatorname{Ad}(Q)J_{r,s}
 &=\omega^{-r}J_{r,s}.
 \label{eq:AdjPQ}
\end{align}
Therefore an adjoint field component $X_{r,s}$ has momenta
\begin{align}
 p_1&=\frac{2\pi n_1}{L_1},
 &
 p_2&=\frac{2\pi n_2}{L_2},
 \\
 p_3&=\frac{2\pi}{L_3}
 \left(n_3+\frac{s}{N}\right),
 &
 p_4&=\frac{2\pi}{L_4}
 \left(n_4-\frac{r}{N}\right),
 \label{eq:SU_N_momenta}
\end{align}
where $n_\mu\in\mathbb{Z}$ and the fractional parts are understood modulo one.
No adjoint component has vanishing shifts in both the 3 and 4 directions.

It follows immediately that the free twisted Faddeev--Popov operator has the
strict lower bound
\begin{equation}
 \lambda_{\min}\bigl(\mathcal{M}_B[0]\bigr)
 =\left(\frac{2\pi}{N}\right)^2
 \min\!\left(\frac{1}{L_3^2},\frac{1}{L_4^2}\right).
 \label{eq:FPgap}
\end{equation}
The intersection of the continuous centralizers of $P$ and $Q$ is empty in
the adjoint Lie algebra; only the discrete center survives.  Equation
\eqref{eq:FPgap} is thus not obtained by deleting an accidental zero mode: it
is a genuine twist-induced gap.

\subsection{Exact enlarged-torus representation}
\label{subsec:enlargedtorus}

The shifted lattice admits a useful exact reorganization.  Define
\begin{equation}
 k_3=Nn_3+s,
 \qquad
 k_4=Nn_4-r.
 \label{eq:k34}
\end{equation}
As $(n_3,n_4,r,s)$ vary with $(r,s)\neq(0,0)$, the pair $(k_3,k_4)$ runs
once over all of $\mathbb{Z}^2$ except the sublattice
$(N\mathbb{Z})^2$.  Introduce the enlarged torus
\begin{equation}
 T_N^4
 =T^4(L_1,L_2,NL_3,NL_4).
 \label{eq:TNdef}
\end{equation}
For any regulated spectral function $F$ of the scalar Laplacian, one then has
\begin{equation}
 \Tr_{\mathrm{Adj},B}F(-\partial^2)
 =\Tr_{T_N^4}F(-\partial^2)
 -\Tr_{T^4}F(-\partial^2).
 \label{eq:twistedtraceTN}
\end{equation}
Subtracting the $N^2-1$ untwisted adjoint components gives the even simpler
identity
\begin{equation}
 \Delta_B\Tr F
 =\Tr_{T_N^4}F(-\partial^2)
 -N^2\Tr_{T^4}F(-\partial^2),
 \label{eq:traceidentity}
\end{equation}
where $\Delta_B\Tr$ denotes twisted minus untwisted.  Since
$\vol(T_N^4)=N^2V_4$, all local heat-kernel coefficients cancel in
Eq.~\eqref{eq:traceidentity}.  The twist dependence is therefore purely
global.

Equation~\eqref{eq:traceidentity} is equivalent to the winding projector
derived in Sec.~\ref{sec:rgz}, but it is often more convenient for numerical
work and for treating zero modes.  In particular, for a massive scalar factor
it gives
\begin{equation}
 \mathcal{D}_N(\mathfrak m)
 =\Ln\Det_{T_N^4}(-\partial^2+\mathfrak m^2)
 -N^2\Ln\Det_{T^4}(-\partial^2+\mathfrak m^2),
 \label{eq:DNtorus}
\end{equation}
with the same common ultraviolet prescription on the two tori.  The formula
also makes transparent why the massless limit requires a separate treatment:
both scalar reference tori contain a geometric zero mode even though the
physical twisted adjoint spectrum does not.

\subsection{The \texorpdfstring{$SU(2)$}{SU(2)} example}

For $N=2$ one may take
\begin{equation}
 \Omega_3=\rmi\sigma_1,
 \qquad
 \Omega_4=\rmi\sigma_2,
 \label{eq:SU2twisteater}
\end{equation}
so that
\begin{equation}
 \Omega_3\Omega_4=-\Omega_4\Omega_3.
 \label{eq:SU2anti}
\end{equation}
In the basis $T^a=\sigma_a/2$,
\begin{align}
 \operatorname{Ad}(\Omega_3)
 &=\operatorname{diag}(+1,-1,-1),
 \\
 \operatorname{Ad}(\Omega_4)
 &=\operatorname{diag}(-1,+1,-1).
 \label{eq:SU2adj}
\end{align}
Thus the three color components obey
\begin{align}
 X^1 &: (\mathrm{P},\mathrm{A}),
 \\
 X^2 &: (\mathrm{A},\mathrm{P}),
 \\
 X^3 &: (\mathrm{A},\mathrm{A})
 \label{eq:SU2bc}
\end{align}
in the $(x_3,x_4)$ directions.  Equivalently, the shifts are
\begin{align}
 (\delta_3^{(1)},\delta_4^{(1)})
 &=\left(0,\frac12\right),
 \\
 (\delta_3^{(2)},\delta_4^{(2)})
 &=\left(\frac12,0\right),
 \\
 (\delta_3^{(3)},\delta_4^{(3)})
 &=\left(\frac12,\frac12\right).
 \label{eq:SU2shifts}
\end{align}
This is the simplest explicit realization of the general spectrum in
Eq.~\eqref{eq:SU_N_momenta}.

\section{Finite-volume RGZ response}
\label{sec:rgz}

\subsection{Transverse RGZ kernel}

In Landau gauge, the tree-level transverse RGZ gluon propagator can be written
as \cite{Dudal:2008sp}
\begin{equation}
 D_{\mu\nu}^{ab}(p)
 =\delta^{ab}
 \left(\delta_{\mu\nu}-\frac{p_\mu p_\nu}{p^2}\right)
 \frac{p^2+M^2}{Q_{\mathrm{RGZ}}(p^2)},
 \label{eq:RGZprop}
\end{equation}
where
\begin{equation}
 Q_{\mathrm{RGZ}}(p^2)
 =p^4+(M^2+m^2)p^2+M^2m^2+\lambda^4,
 \label{eq:QRGZ}
\end{equation}
and
\begin{equation}
 \lambda^4=2g^2N\gamma^4.
 \label{eq:lambda}
\end{equation}
Here $\gamma$ is the Gribov parameter, while $m^2$ and $M^2$ encode the
condensate contributions in the standard RGZ parametrization.  The transverse
inverse kernel is
\begin{equation}
 K_{\mathrm{RGZ}}(p^2)
 =\frac{Q_{\mathrm{RGZ}}(p^2)}{p^2+M^2}.
 \label{eq:KRGZ}
\end{equation}
We factorize the quartic polynomial as
\begin{equation}
 Q_{\mathrm{RGZ}}(p^2)
 =(p^2+\mu_+^2)(p^2+\mu_-^2),
 \label{eq:factorQ}
\end{equation}
with
\begin{equation}
 \mu_\pm^2
 =\frac{M^2+m^2}{2}
 \mathbin{\pm}
 \frac12
 \sqrt{(M^2-m^2)^2-4\lambda^4}.
 \label{eq:mupm2}
\end{equation}
When the two roots are complex, we choose square roots $\mu_\pm$ with positive
real parts.  In the usual complex-conjugate regime,
$\mu_-^2=(\mu_+^2)^*$ and the final determinant difference is real.

At this stage we hold $\gamma$, $m^2$, and $M^2$ fixed and compute only the
transverse spectral contribution.  The sector dependence of these parameters
and the completion of the full measure are deferred to
Sec.~\ref{sec:completion}.

\subsection{A universal shifted determinant}

For a scalar spectral factor $p^2+\mathfrak m^2$, define the color-summed
minimal-twist difference
\begin{align}
 \mathcal{D}_N(\mathfrak m)
 ={}&
 \sum_{(r,s)\neq(0,0)}
 \sum_{\bm n\in\mathbb{Z}^4}
 \ln\!\left[
 \bigl(p_{r,s}(\bm n)\bigr)^2+\mathfrak m^2
 \right]
 \nonumber\\
 &-(N^2-1)
 \sum_{\bm n\in\mathbb{Z}^4}
 \ln\!\left[p_0(\bm n)^2+\mathfrak m^2\right],
 \label{eq:DNdef}
\end{align}
where $p_{r,s}$ is given by Eq.~\eqref{eq:SU_N_momenta} and
\begin{equation}
 p_{0,\mu}(\bm n)=\frac{2\pi n_\mu}{L_\mu}.
 \label{eq:p0}
\end{equation}
For $\re\mathfrak m>0$, the difference can be evaluated without choosing a
separate ultraviolet subtraction for the two sectors.

Using
\begin{equation}
 \ln A
 =-\int_0^\infty\frac{\rmd t}{t}\,
 \rme^{-tA}
 \label{eq:Schwinger}
\end{equation}
inside a difference, Poisson resummation gives
\begin{equation}
 \mathcal{D}_N(\mathfrak m)
 =-\frac{V_4\mathfrak m^2}{2\pi^2}
 \sum_{\bm\ell\in\mathbb{Z}^4\setminus\{0\}}
 \mathcal{C}_N(\ell_3,\ell_4)
 \frac{K_2(\mathfrak m\rho_{\bm\ell})}
 {\rho_{\bm\ell}^2},
 \label{eq:DNbeforecolor}
\end{equation}
where
\begin{align}
 V_4&=L_1L_2L_3L_4,
 \\
 \rho_{\bm\ell}^2
 &=\sum_{\mu=1}^4 L_\mu^2\ell_\mu^2,
 \label{eq:rho}
\end{align}
and the color factor is
\begin{align}
 \mathcal{C}_N(\ell_3,\ell_4)
 ={}&
 \sum_{(r,s)\neq(0,0)}
 \left[
 \exp\!\left(
 \frac{2\pi\rmi}{N}
 (\ell_3s-\ell_4r)
 \right)-1
 \right].
 \label{eq:CNdef}
\end{align}
The finite-dimensional sum can be done exactly:
\begin{equation}
 \mathcal{C}_N(\ell_3,\ell_4)
 =N^2
 \left[
 \delta_{\ell_3=0\ (\mathrm{mod}\ N)}
 \delta_{\ell_4=0\ (\mathrm{mod}\ N)}-1
 \right].
 \label{eq:CNresult}
\end{equation}
Thus
\begin{equation}
 \mathcal{C}_N(\ell_3,\ell_4)
 =\begin{cases}
 0,
 &\text{$\ell_3$, $\ell_4\in N\mathbb{Z}$} ,
 \\[2mm]
 -N^2,
 &\text{otherwise}.
 \end{cases}
 \label{eq:CNcases}
\end{equation}
Substitution into Eq.~\eqref{eq:DNbeforecolor} yields our basic finite-volume
formula,
\begin{equation}
 \mathcal{D}_N(\mathfrak m)
 =\frac{N^2V_4\mathfrak m^2}{2\pi^2}
 \sum_{\substack{\bm\ell\in\mathbb{Z}^4\\
 (\ell_3,\ell_4)\notin(N\mathbb{Z})^2}}
 \frac{K_2(\mathfrak m\rho_{\bm\ell})}
 {\rho_{\bm\ell}^2}.
 \label{eq:DNfinal}
\end{equation}
For $N=2$, Eq.~\eqref{eq:CNcases} is the statement that the three adjoint
phases sum to zero for even-even windings and give $-4$ for every other
$(\ell_3,\ell_4)$ parity class.

Several useful properties are manifest in Eq.~\eqref{eq:DNfinal}.  The local
zero-winding heat-kernel coefficient cancels between the two sectors, so the
ultraviolet divergence is absent from the difference.  The response is also
nonlocal in the expected finite-volume sense: it is expressed entirely in
terms of worldline windings around the torus.  The center twist enters only as
an arithmetic projector on those windings.

\subsection{Exact transverse RGZ determinant difference}

The transverse quadratic contribution in $d$ Euclidean dimensions is
\begin{equation}
 \Gamma_{T,B}^{(2)}
 =\frac{d-1}{2}
 \sum_{\text{adj. modes}}
 \ln K_{\mathrm{RGZ}}(p^2).
 \label{eq:GammaTB}
\end{equation}
Using Eqs.~\eqref{eq:KRGZ} and \eqref{eq:factorQ}, the twisted-minus-untwisted
difference on $T^4$ is
\begin{equation}
 \Delta\Gamma_T^{(2)}(B)
 =\frac32
 \left[
 \mathcal{D}_N(\mu_+)
 +\mathcal{D}_N(\mu_-)
 -\mathcal{D}_N(M)
 \right].
 \label{eq:DeltaGammaT_D}
\end{equation}
Combining this with Eq.~\eqref{eq:DNfinal}, we obtain
\begin{align}
 \Delta\Gamma_T^{(2)}(B)
 ={}&\frac{3N^2V_4}{4\pi^2}
 \sum_{\substack{\bm\ell\in\mathbb{Z}^4\\
 (\ell_3,\ell_4)\notin(N\mathbb{Z})^2}}
 \frac{1}{\rho_{\bm\ell}^2}
 \Bigl[
 \mu_+^2K_2(\mu_+\rho_{\bm\ell})
 \nonumber\\
 &\hspace{32mm}
 +\mu_-^2K_2(\mu_-\rho_{\bm\ell})
 -M^2K_2(M\rho_{\bm\ell})
 \Bigr].
 \label{eq:DeltaGammaTfinal}
\end{align}
Equation~\eqref{eq:DeltaGammaTfinal} is the main analytic result of this
paper.  It is valid for fixed RGZ parameters in the regime in which the
screening masses appearing in the Bessel representation have positive real
parts.  Analytic continuation from positive real masses defines the
complex-conjugate-pole case.

It is useful to define the associated transverse ratio
\begin{equation}
 R_T^{(2)}[B]
 =\exp\left\{-\Delta\Gamma_T^{(2)}(B)\right\}.
 \label{eq:RTdef}
\end{equation}
We stress that $R_T^{(2)}$ is not yet the physical ratio in
Eq.~\eqref{eq:Rdef}; it is one controlled component of the RGZ finite-volume
response.

\subsection{Large-volume behavior and complex poles}
\label{subsec:largeL}

Consider the uniform scaling
\begin{equation}
 L_\mu=L\widehat L_\mu,
 \qquad
 L\longrightarrow\infty,
 \label{eq:uniformscale}
\end{equation}
with fixed aspect ratios $\widehat L_\mu$.  For $\re z>0$,
\begin{equation}
 K_2(z)
 =\sqrt{\frac{\pi}{2z}}\,
 \rme^{-z}
 \left[1+\mathcal{O}(z^{-1})\right].
 \label{eq:K2asymptotic}
\end{equation}
Therefore every term in Eq.~\eqref{eq:DeltaGammaTfinal} is exponentially
small.  If
\begin{equation}
 m_*
 =\min\{M,\re\mu_+,\re\mu_-\}>0,
 \label{eq:mstar}
\end{equation}
then, up to an aspect-ratio-dependent coefficient,
\begin{equation}
 \Delta\Gamma_T^{(2)}(B)
 =\mathcal{O}\!\left(
 L^{3/2}\rme^{-m_*L\rho_*}
 \right),
 \label{eq:largeLgeneral}
\end{equation}
where $\rho_*$ is the shortest allowed dimensionless winding length.
Consequently,
\begin{equation}
 R_T^{(2)}[B]\longrightarrow1.
 \label{eq:RTstrong}
\end{equation}
Within this massive transverse Gaussian sector, the minimal center twist is
therefore forgotten exponentially fast in the thermodynamic limit.  This is
precisely the limiting behavior required by the strong center-vortex criterion
\eqref{eq:strongcriterion}, although the full gauge-fixed measure must still be
included before drawing a physical conclusion.

For an isotropic torus, $L_\mu=L$, the shortest allowed windings are
$(\ell_3,\ell_4)=(\pm1,0)$ and $(0,\pm1)$ with $\ell_1=\ell_2=0$.  The leading
term is
\begin{align}
 \Delta\Gamma_T^{(2)}(B)
 \simeq{}&\frac{3N^2L^2}{\pi^2}
 \Bigl[
 \mu_+^2K_2(\mu_+L)
 +\mu_-^2K_2(\mu_-L)
 \nonumber\\
 &\hspace{34mm}
 -M^2K_2(ML)
 \Bigr]
 \label{eq:leadingisotropic}
\end{align}
up to contributions suppressed by longer windings.
If $\mu_-=(\mu_+)^*$, the first two terms combine into
\begin{equation}
 2\re\!\left[
 \mu_+^2K_2(\mu_+L)
 \right].
 \label{eq:complexpair}
\end{equation}
Their large-$L$ dependence is a decaying oscillation whenever
$\im\mu_+\neq0$.  Whether this oscillatory component is leading depends on the
relative size of $M$ and $\re\mu_+$.  This offers a simple finite-volume
spectral signature of the complex RGZ pole structure without assigning a
particle interpretation to those poles.

The massless limits of Eq.~\eqref{eq:DNfinal} are nonuniform because the
untwisted sector contains zero modes and the large-proper-time part of the
heat kernel changes qualitatively.  We therefore do not use
Eq.~\eqref{eq:DeltaGammaTfinal} to infer a perturbative massless result by
simply setting the RGZ scales to zero.

\section{Gaussian completion of the twisted RGZ partition function}
\label{sec:completion}

\subsection{Quadratic measure and cancellation of the localizing determinants}
\label{subsec:gaussianmeasure}

We now complete the quadratic functional integral at fixed RGZ parameters.
For the determinant counting it is cleaner to impose Landau gauge exactly
rather than introduce a finite-$\alpha$ extension of the localized action.
Let
\begin{equation}
 z=p^2>0
 \label{eq:zpositive}
\end{equation}
be a nonzero eigenvalue of the adjoint scalar Laplacian in the chosen sector.
After the Zwanziger fields are integrated out, the transverse gluon kernel is
\begin{equation}
 K_T(z)
 =z+m^2+\frac{\lambda^4}{z+M^2}
 =\frac{Q_{\mathrm{RGZ}}(z)}{z+M^2},
 \label{eq:KTfull}
\end{equation}
which is the kernel already used in Eq.~\eqref{eq:KRGZ}.

The cancellation of the localizing determinants can be made explicit without
suppressing their multiplicity.  Write $d_G=N^2-1$ and let
\begin{equation}
 H_B^{ab}
 =\bigl(-\partial_B^2+M^2\bigr)^{ab}
 \label{eq:HB}
\end{equation}
act on the first adjoint index of
$\varphi_\mu^{ac}$, $\bar\varphi_\mu^{ac}$, $\omega_\mu^{ac}$, $\bar\omega_\mu^{ac}$; the Lorentz index $\mu$ and the second
adjoint index $c$ are spectators.  There are therefore $d\times d_G$ identical
copies of $H_B$.  Mode by mode, the bosonic and Grassmann normalization
factors are
\begin{equation}
 \prod_{\mu=1}^{d}\prod_{c=1}^{d_G}
 (\Det H_B)^{-1}(\Det H_B)=1.
 \label{eq:auxmultiplicity}
\end{equation}
The $A$--$\varphi$ mixing acts as a source for the bosonic field.  Suppressing
only the spectator labels, the remaining Gaussian identity is
\begin{align}
 &\int \rmd\varphi\,\rmd\bar\varphi\,
       \rmd\omega\,\rmd\bar\omega\,
 \exp\bigl[
 -\bar\varphi H_B\varphi
 -\bar\omega H_B\omega
 +\bar J\varphi+\bar\varphi J
 \bigr]
 \nonumber\\
 &\hspace{25mm}\propto
 \frac{\Det H_B}{\Det H_B}
 \exp\bigl(\bar JH_B^{-1}J\bigr).
 \label{eq:auxcancel}
\end{align}
Together with $f^{abc}f^{dbc}=N\delta^{ad}$ and the standard RGZ sign
conventions, the source term reproduces the horizon part of the transverse
kernel, $\lambda^4/(z+M^2)$ with
$\lambda^4=2g^2N\gamma^4$ 
\cite{Dudal:2008sp,Dudal:2010tf}.  No additional massive oscillator
determinant remains after the localizing fields are integrated out.

Equations~\eqref{eq:auxmultiplicity} and \eqref{eq:auxcancel} assume that the
two localizing systems are globally patched so that they represent the same
operator $H_B$.  This is not an innocuous assumption on a nontrivial bundle.
We take the globally defined nonlocal horizon functional as primary and
require any localization to reproduce it.  Wilson-line dressings provide one
model for implementing this requirement in a background-gauge-covariant
construction \cite{Kroff:2018}.

It remains to combine the gluon constraint with the Faddeev--Popov
determinant.  For a normalized nonzero scalar eigenmode, decompose its vector
amplitude as
\begin{equation}
 A_\mu=A_\mu^T+\frac{p_\mu}{\sqrt z}\,a_L,
 \qquad
 p_\mu A_\mu^T=0.
 \label{eq:modeTLdecomp}
\end{equation}
Then $p_\mu A_\mu=\sqrt z\,a_L$, and the exact Landau-gauge delta function
gives
\begin{equation}
 \delta(p_\mu A_\mu)
 =z^{-1/2}\delta(a_L).
 \label{eq:landaudeltajac}
\end{equation}
The Faddeev--Popov ghost pair contributes the eigenvalue $z$
on the same mode.  Consequently the longitudinal amplitude, the gauge-fixing delta
function, and the ghost determinant combine into $z^{1/2}$, whereas the
$d-1$ transverse real amplitudes give $K_T(z)^{-(d-1)/2}$.  Thus
\begin{equation}
 Z_p^{(2)}
 \propto
 K_T(z)^{-(d-1)/2}z^{1/2},
 \label{eq:Zpermodefull}
\end{equation}
and the nonzero-mode one-loop effective action is
\begin{equation}
 \Gamma_p^{(2)}
 =\frac{d-1}{2}\ln K_T(z)
 -\frac12\ln z.
 \label{eq:Gammapermodefull}
\end{equation}
This derivation does not require assigning a horizon term to a longitudinal
mode away from Landau gauge.  The second term is the finite-volume remnant of
the gauge-fixing/ghost sector.  On $\mathbb{R}^d$ it is a scaleless
determinant in dimensional regularization, which explains why the standard
one-loop RGZ vacuum energy is reported solely in terms of the massive
transverse kernel \cite{Dudal:2008sp}.  On a torus it cannot be discarded.

\subsection{Complete nonzero-mode determinant}
\label{subsec:nonzerocomplete}

To treat the massless factor without ambiguity, introduce a renormalization
scale $\mu_{\mathrm R}$ and define the primed massive determinant difference
\begin{align}
 \widehat{\mathcal D}_N(\mathfrak m;\mu_{\mathrm R})
 ={}&
 \Ln\Det'_{T_N^4}
 \frac{-\partial^2+\mathfrak m^2}{\mu_{\mathrm R}^2}
 \nonumber\\
 &-N^2\Ln\Det'_{T^4}
 \frac{-\partial^2+\mathfrak m^2}{\mu_{\mathrm R}^2}.
 \label{eq:Dhatdef}
\end{align}
The prime removes the geometric zero mode of each scalar reference torus.  For
$\mathfrak m\neq0$, comparison with the unprimed determinant in
Eq.~\eqref{eq:DNtorus} gives
\begin{equation}
 \widehat{\mathcal D}_N(\mathfrak m;\mu_{\mathrm R})
 =\mathcal D_N(\mathfrak m)
 +(N^2-1)\ln\frac{\mathfrak m^2}{\mu_{\mathrm R}^2}.
 \label{eq:Dhatrelation}
\end{equation}
We also define the massless gauge-shape determinant
\begin{equation}
 \mathcal G_N(\mu_{\mathrm R})
 =\widehat{\mathcal D}_N(0;\mu_{\mathrm R}).
 \label{eq:GNdef}
\end{equation}
No new spectral calculation is needed to obtain this quantity.  The primed
massive determinant has a smooth $\mathfrak m\to0$ limit, so
Eq.~\eqref{eq:Dhatrelation} implies
\begin{equation}
 \mathcal G_N(\mu_{\mathrm R})
 =\lim_{\mathfrak m\to0^+}
 \left[
 \mathcal D_N(\mathfrak m)
 +(N^2-1)\ln\frac{\mathfrak m^2}{\mu_{\mathrm R}^2}
 \right].
 \label{eq:GNmasslesslimit}
\end{equation}
Thus the same Bessel representation in Eq.~\eqref{eq:DNfinal}, supplemented
by the explicit logarithmic subtraction in Eq.~\eqref{eq:GNmasslesslimit},
determines the massless gauge factor as well.

Combining Eq.~\eqref{eq:Gammapermodefull} with the enlarged-torus identity,
the complete twisted-minus-untwisted contribution of all \textit{nonzero}
modes is
\begin{align}
 \Delta\Gamma_{\mathrm{nz}}^{(2)}(B)
 ={}&\frac{d-1}{2}
 \Bigl[
 \widehat{\mathcal D}_N(\mu_+;\mu_{\mathrm R})
 +\widehat{\mathcal D}_N(\mu_-;\mu_{\mathrm R})
 \nonumber\\
 &\hspace{28mm}
 -\widehat{\mathcal D}_N(M;\mu_{\mathrm R})
 \Bigr]
 -\frac12\mathcal G_N(\mu_{\mathrm R}).
 \label{eq:DeltaGammaznonzero}
\end{align}
This is an exact Gaussian one-loop spectral result at fixed RGZ parameters.
No further oscillator determinant is missing.  The first line is the massive
RGZ sector and the final term is the universal massless gauge factor.

A convenient zeta-function definition of the latter is obtained from the
scalar heat kernels $\Theta_{T}(t)=\Tr_T \rme^{t\partial^2}$.  Set
\begin{align}
 \zeta_{\mathcal G}(s)
 ={}&\frac{\mu_{\mathrm R}^{2s}}{\Gamma(s)}
 \int_0^\infty \rmd t\,t^{s-1}
 \Bigl[
 \Theta_{T_N^4}(t)-1
 \nonumber\\
 &\hspace{26mm}
 -N^2\bigl(\Theta_{T^4}(t)-1\bigr)
 \Bigr],
 \label{eq:zetaG}
\end{align}
where analytic continuation to $s=0$ is understood.  Then
\begin{equation}
 \mathcal G_N(\mu_{\mathrm R})
 =-\zeta_{\mathcal G}'(0).
 \label{eq:GNzeta}
\end{equation}
Equivalently, if
\begin{equation}
 E_4(s;L_1,L_2,L_3,L_4)
 =\sum_{\bm n\in\mathbb Z^4\setminus\{0\}}
 \left(
 \sum_{\mu=1}^4\frac{n_\mu^2}{L_\mu^2}
 \right)^{-s}
 \label{eq:Epstein4}
\end{equation}
denotes the anisotropic Epstein zeta function, then
\begin{align}
 \mathcal G_N(\mu_{\mathrm R})
 ={}&-\left.\frac{\rmd}{\rmd s}\right|_{s=0}
 \left(\frac{\mu_{\mathrm R}}{2\pi}\right)^{2s}
 \Bigl[
 E_4(s;L_1,L_2,NL_3,NL_4)
 \nonumber\\
 &\hspace{30mm}
 -N^2E_4(s;L_1,L_2,L_3,L_4)
 \Bigr].
 \label{eq:GNEpstein}
\end{align}
The zero-winding ultraviolet terms cancel because
$\vol(T_N^4)=N^2V_4$, while subtraction of the two scalar zero modes controls
the large-$t$ limit.

For a flat torus, the primed scalar Laplacian has $\zeta(0)=-1$.  Under a
uniform rescaling $L_\mu\to aL_\mu$ one therefore finds
\begin{equation}
 \mathcal G_N(aL_\mu;\mu_{\mathrm R})
 =\mathcal G_N(L_\mu;\mu_{\mathrm R})
 -2(N^2-1)\ln a.
 \label{eq:GNscaling}
\end{equation}
Unlike the massive Bessel sum, the massless gauge factor is not exponentially
small.  This scaling will be important when discussing global zero modes.

\subsection{Constant gluons and the global stabilizer}
\label{subsec:zeromodes}

The irreducibly twisted adjoint spectrum has no $p=0$ mode.  The untwisted
sector does.  The ghost zero modes generate the residual global
$SU(N)$ gauge transformations and must be omitted from
$\Det\mathcal M$; this is the origin of the primes above.  Constant gluons,
on the other hand, have a finite quadratic RGZ kernel
\begin{align}
 K_0
 &=K_T(0)
 =m^2+\frac{\lambda^4}{M^2}
 \nonumber\\
 &=\frac{M^2m^2+\lambda^4}{M^2}
 =\frac{\mu_+^2\mu_-^2}{M^2}.
 \label{eq:Kzero}
\end{align}
For convergence of the original real Gaussian integrals, we require
$M^2>0$ and $K_T(z)>0$ on the nonnegative real spectrum, in particular
$K_0>0$.  These conditions are compatible with a complex-conjugate pair of
factorization masses because $Q_{\mathrm{RGZ}}(z)$ itself remains real for
real $z$.
At the Gaussian level the $d(N^2-1)$ real constant gluon components of the
untwisted sector therefore contribute
\begin{equation}
 \Gamma_{A,0}^{(2)}
 =\frac{d}{2}(N^2-1)
 \ln\frac{K_0}{\mu_{\mathrm R}^2}.
 \label{eq:GammaAzero}
\end{equation}
Because our convention is twisted minus untwisted, this term enters the
sector difference with a minus sign.

Using Eq.~\eqref{eq:Dhatrelation}, the nonzero-mode expression
Eq.~\eqref{eq:DeltaGammaznonzero} plus Eq.~\eqref{eq:GammaAzero} collapses to
a particularly simple formula in terms of the transverse result already
derived in Sec.~\ref{sec:rgz}:
\begin{align}
 \Delta\Gamma_{\mathrm{Gauss}}^{(2)}(B)
 ={}&\Delta\Gamma_T^{(2)}(B)
 -\frac12\mathcal G_N(\mu_{\mathrm R})
 \nonumber\\
 &-\frac{N^2-1}{2}
 \ln\frac{K_0}{\mu_{\mathrm R}^2}
 +\Delta\Gamma_{\mathrm{stab}}.
 \label{eq:masterGaussian}
\end{align}
Equation~\eqref{eq:masterGaussian} is the Gaussian completion of
Eq.~\eqref{eq:DeltaGammaTfinal}.  The first term is the exponentially
screened massive RGZ response, the second is the massless finite-volume gauge
factor, and the third corrects the counting of the $p=0$ gluon mode.

The last term, $\Delta\Gamma_{\mathrm{stab}}$, is qualitatively different.
It is not the determinant of another local differential operator.  Let $\mathcal H_B$
denote the stabilizer subgroup of the flat representative in sector $B$, and
let $J_B$ denote the finite-dimensional Jacobian obtained when the associated
zero modes are extracted with a common microscopic measure.  The residual
factor can be parameterized as
\begin{equation}
 Z_{\mathrm{stab},B}
 =\frac{J_B}{\vol_\mu(\mathcal H_B)},
 \qquad
 \Delta\Gamma_{\mathrm{stab}}
 =\ln\frac{\vol_\mu(\mathcal H_B)}{\vol_\mu(\mathcal H_0)}
 -\ln\frac{J_B}{J_0}.
 \label{eq:stabilizermeasure}
\end{equation}
Here $\vol_\mu$ is the volume induced by the chosen functional measure, not
an independently normalized abstract Haar volume. Equation~\eqref{eq:stabilizermeasure} is a parametrization of the finite-dimensional
zero-mode measure, rather than an additional continuum prediction;
its absolute value requires a common microscopic normalization.  For the representatives
used here, $\mathcal H_0=SU(N)$ and $\mathcal H_B=\mathbb Z_N$; if one quotients from
the outset by gauge transformations acting trivially on adjoint fields, the
equivalent statement is $\mathcal H_0=PSU(N)$ and $\mathcal H_B$ trivial.  A Gaussian
Hessian with primed ghost determinant fixes neither $J_B/J_0$ nor the induced
relative group volume.  In particular, $\Delta\Gamma_{\mathrm{stab}}$ should
not be treated as a volume-independent numerical constant.  A common lattice
Haar measure, or a continuum prescription that fixes the residual global
gauge symmetry and the normalization of its zero modes in both sectors, fixes
Eq.~\eqref{eq:stabilizermeasure}.

The determinant calculation fixes two nontrivial consistency conditions on
this residual measure.  First, changing the determinant scale gives
\begin{equation}
 \mathcal G_N(L_\mu;b\mu_{\mathrm R})
 =\mathcal G_N(L_\mu;\mu_{\mathrm R})
 -2(N^2-1)\ln b.
 \label{eq:GNmuscaling}
\end{equation}
Define the part fixed by the oscillator and constant-gluon integrations by
\begin{equation}
 \Delta\Gamma_{\mathrm{osc}}
 =\Delta\Gamma_{\mathrm{Gauss}}^{(2)}
 -\Delta\Gamma_{\mathrm{stab}}.
 \label{eq:Gammaoscdef}
\end{equation}
At fixed RGZ parameters, Eq.~\eqref{eq:masterGaussian} then implies
\begin{equation}
 \Delta\Gamma_{\mathrm{osc}}(b\mu_{\mathrm R})
 -\Delta\Gamma_{\mathrm{osc}}(\mu_{\mathrm R})
 =2(N^2-1)\ln b.
 \label{eq:Gammaoscmuscaling}
\end{equation}
The stabilizer contribution must have the opposite dependence so that the
physical ratio is independent of the arbitrary scale $\mu_{\mathrm R}$.
Second, under a uniform rescaling $L_\mu\to aL_\mu$ with $a\to\infty$,
Eqs.~\eqref{eq:GNscaling} and \eqref{eq:largeLgeneral} give
\begin{align}
 \Delta\Gamma_{\mathrm{osc}}(aL_\mu)
 ={}&(N^2-1)\ln a+C_{\mathrm{osc}}
 \nonumber\\
 &+\mathcal{O}\!\left(
 a^{3/2}\rme^{-a m_*\rho_{\min}}
 \right),
 \label{eq:GammaoscLscaling}
\end{align}
where
\begin{equation}
 C_{\mathrm{osc}}
 =-\frac12\mathcal G_N(L_\mu;\mu_{\mathrm R})
 -\frac{N^2-1}{2}\ln\frac{K_0}{\mu_{\mathrm R}^2}
 \label{eq:Coscref}
\end{equation}
and $\rho_{\min}$ is the shortest allowed winding length before the
rescaling.  If the fully normalized ratio has a nonzero thermodynamic limit,
the zero-mode/stabilizer measure must in particular cancel the displayed
$(N^2-1)\ln a$.  Equations~\eqref{eq:Gammaoscmuscaling} and
\eqref{eq:GammaoscLscaling} are quantitative matching conditions for a
lattice or other microscopic normalization.

This point changes the interpretation of the large-volume limit.  The massive
term in Eq.~\eqref{eq:masterGaussian} tends to zero exponentially, but
Eq.~\eqref{eq:GNscaling} contains a logarithmic size dependence.  It would be
incorrect to interpret that logarithm alone as a physical violation of the
strong center-vortex criterion: its coefficient is tied to the mismatch of
global zero-mode sectors and must be combined with the stabilizer measure.
The Gaussian calculation has thus isolated, rather than hidden, the precise
normalization problem.

\subsection{Sectorwise stationary conditions and an explicit horizon source}
\label{subsec:gapequations}

The preceding formulas kept the RGZ parameters fixed.  Let
\begin{equation}
 x_i=(\lambda^4,m^2,M^2,\ldots)
 \label{eq:xi}
\end{equation}
and let $U_{\mathrm{RGZ}}(x_i)$ denote the renormalized local vacuum potential,
including the horizon and local-composite-operator vacuum terms appropriate
to the chosen RGZ scheme.  At Gaussian order the sectorwise effective action
has the structure
\begin{equation}
 \Gamma_B(x_i)
 =V_4U_{\mathrm{RGZ}}(x_i)
 +\Gamma_{\mathrm{spec},B}^{(2)}(x_i)
 +\Gamma_{\mathrm{stab},B}
 +\cdots,
 \label{eq:GammaBfull}
\end{equation}
where the ellipsis denotes interactions beyond Gaussian order.  The
stationary conditions are
\begin{equation}
 \frac{\partial\Gamma_B}{\partial x_i}=0.
 \label{eq:gapeqgeneric}
\end{equation}
The local term is the same function of $x_i$ in the two flux sectors; their
finite-volume difference enters through the spectral sums and through the
global normalization.

The massless determinant $\mathcal G_N$ is independent of the RGZ parameters.
If the stabilizer normalization is also chosen independently of $x_i$, the
Gaussian sector-difference source obeys
\begin{equation}
 \frac{\partial\Delta\Gamma_{\mathrm{Gauss}}^{(2)}}{\partial x_i}
 =\frac{\partial\Delta\Gamma_T^{(2)}}{\partial x_i}
 -\frac{N^2-1}{2}
 \frac{\partial\ln K_0}{\partial x_i}.
 \label{eq:gapmasterderivative}
\end{equation}
Thus the Bessel sum derived previously already contains the complete
\textit{extensive spectral} source for the Gaussian gap equations; the extra
zero-mode term is only of order unity and hence of order $1/V_4$ in the
effective potential.

For the horizon parameter this source can be written in closed form.  Define
the shifted resolvent difference
\begin{equation}
 \mathcal R_N(\mathfrak m)
 =\frac{\partial\mathcal D_N(\mathfrak m)}
 {\partial\mathfrak m^2}.
 \label{eq:RNdef}
\end{equation}
Differentiating Eq.~\eqref{eq:DNfinal} and using the Bessel identity
$\rmd[z^2K_2(z)]/\rmd z=-z^2K_1(z)$ gives
\begin{equation}
 \mathcal R_N(\mathfrak m)
 =-\frac{N^2V_4\mathfrak m}{4\pi^2}
 \sum_{\substack{\bm\ell\in\mathbb{Z}^4\\
 (\ell_3,\ell_4)\notin(N\mathbb{Z})^2}}
 \frac{K_1(\mathfrak m\rho_{\bm\ell})}
 {\rho_{\bm\ell}}.
 \label{eq:RNbessel}
\end{equation}
Since
\begin{equation}
 \frac{1}{Q_{\mathrm{RGZ}}(z)}
 =\frac{1}{\mu_-^2-\mu_+^2}
 \left[
 \frac{1}{z+\mu_+^2}
 -\frac{1}{z+\mu_-^2}
 \right],
 \label{eq:Qpartialfraction}
\end{equation}
we obtain
\begin{equation}
 \frac{\partial\Delta\Gamma_T^{(2)}}{\partial\lambda^4}
 =\frac{d-1}{2}
 \frac{\mathcal R_N(\mu_+)-\mathcal R_N(\mu_-)}
 {\mu_-^2-\mu_+^2}.
 \label{eq:dGammaTdLambda}
\end{equation}
Combining Eqs.~\eqref{eq:gapmasterderivative} and \eqref{eq:Kzero} yields the
finite-volume correction to the horizon equation,
\begin{align}
 \frac{\partial\Delta\Gamma_{\mathrm{Gauss}}^{(2)}}
 {\partial\lambda^4}
 ={}&\frac{d-1}{2}
 \frac{\mathcal R_N(\mu_+)-\mathcal R_N(\mu_-)}
 {\mu_-^2-\mu_+^2}
 \nonumber\\
 &-\frac{N^2-1}
 {2(M^2m^2+\lambda^4)}.
 \label{eq:dGammaFullLambda}
\end{align}
This is directly usable once a renormalized RGZ vacuum potential is chosen.
The other two derivatives can also be closed analytically.  Define
\begin{equation}
 \sigma
 =\sqrt{(M^2-m^2)^2-4\lambda^4}
 =\mu_+^2-\mu_-^2,
 \qquad
 \delta=M^2-m^2,
 \label{eq:sigmadelta}
\end{equation}
with the same branch convention as in Eq.~\eqref{eq:mupm2}.  Differentiating
the two roots gives
\begin{align}
 \frac{\partial\mu_+^2}{\partial m^2}
 &=\frac12-\frac{\delta}{2\sigma},
 &
 \frac{\partial\mu_-^2}{\partial m^2}
 &=\frac12+\frac{\delta}{2\sigma},
 \nonumber\\
 \frac{\partial\mu_+^2}{\partial M^2}
 &=\frac12+\frac{\delta}{2\sigma},
 &
 \frac{\partial\mu_-^2}{\partial M^2}
 &=\frac12-\frac{\delta}{2\sigma}.
 \label{eq:rootderivatives}
\end{align}
Using Eq.~\eqref{eq:gapmasterderivative} and
$K_0=(M^2m^2+\lambda^4)/M^2$, one obtains
\begin{align}
 \frac{\partial\Delta\Gamma_{\mathrm{Gauss}}^{(2)}}{\partial m^2}
 ={}&\frac{d-1}{4}
 \left[
 \mathcal R_N(\mu_+)+\mathcal R_N(\mu_-)
 +\frac{\delta}{\sigma}
 \bigl(\mathcal R_N(\mu_-)-\mathcal R_N(\mu_+)\bigr)
 \right]
 \nonumber\\
 &-\frac{(N^2-1)M^2}{2(M^2m^2+\lambda^4)},
 \label{eq:dGammaFullm}
\\[1mm]
 \frac{\partial\Delta\Gamma_{\mathrm{Gauss}}^{(2)}}{\partial M^2}
 ={}&\frac{d-1}{4}
 \left[
 \mathcal R_N(\mu_+)+\mathcal R_N(\mu_-)
 +\frac{\delta}{\sigma}
 \bigl(\mathcal R_N(\mu_+)-\mathcal R_N(\mu_-)\bigr)
 -2\mathcal R_N(M)
 \right]
 \nonumber\\
 &+\frac{(N^2-1)\lambda^4}
 {2M^2(M^2m^2+\lambda^4)}.
 \label{eq:dGammaFullM}
\end{align}
Equations~\eqref{eq:dGammaFullLambda}, \eqref{eq:dGammaFullm}, and
\eqref{eq:dGammaFullM} are the complete fixed-parameter Gaussian sources for
the three RGZ mass parameters.  Their apparent $1/\sigma$ singularities at a
double root are removable; the coincident-root value is obtained by taking
the continuous derivative limit of $\mathcal R_N$.

For later comparison with an untwisted stationary point $x_i^{(0)}$, define
the Hessian of the untwisted effective potential by
\begin{equation}
 H_{ij}
 =\left.
 \frac{\partial^2\mathcal V_0}
 {\partial x_i\partial x_j}
 \right|_{x=x^{(0)}}.
 \label{eq:Hessian}
\end{equation}
If it is nonsingular, the leading finite-volume shift is
\begin{equation}
 \sum_jH_{ij}\,\delta x_j
 =-\left.
 \frac{\partial\Delta\mathcal V_B}{\partial x_i}
 \right|_{x=x^{(0)}}.
 \label{eq:linearshift}
\end{equation}
Equations~\eqref{eq:RNbessel} and
\eqref{eq:dGammaFullLambda}--\eqref{eq:dGammaFullM} provide the explicit
source needed for this calculation.

\subsection{What remains beyond the Gaussian spectral calculation}
\label{subsec:beyondgaussian}

Equation~\eqref{eq:masterGaussian} substantially narrows the unfinished part
of the problem.  At fixed RGZ parameters, all nonzero quadratic oscillator
determinants and the constant gluon Gaussian are explicit.  Three issues
remain before identifying the result with the physical twisted ratio
$Z[B]/Z[0]$.

First, the finite-dimensional zero-mode/stabilizer factor in
Eq.~\eqref{eq:stabilizermeasure} must be fixed by a common microscopic
normalization.  Second, if $m^2$, $M^2$, and $\lambda^4$ are to be determined
dynamically, one must choose a definite renormalized RGZ/LCO effective
potential and solve its sectorwise stationary equations.  Third, interactions
beyond the Gaussian approximation can modify both the finite-volume response
and the interpretation of the massless gauge-fixing/ghost sector.  Background-field and
BRST-invariant RGZ formulations are natural frameworks in which to test the
representative and gauge-parameter independence of this completion
\cite{Capri:2016aqq,Justo:2022eyg,Dudal:2023rki,Kroff:2018}.

The distinction is important: the missing object is no longer an unspecified
``full RGZ determinant.''  The spectral determinant is known at Gaussian
order. \revone{Within the fixed-representative Gaussian calculation, the
remaining normalization ambiguity is finite-dimensional and
global.
The dynamical determination of the RGZ vacuum and corrections
beyond Gaussian order are separate issues.}
%the remaining ambiguity is finite-dimensional and global, together with the genuinely dynamical choice of the RGZ vacuum and higher-loop corrections.

\section{From a global twist to a dynamical center vortex}
\label{sec:vortex}

\subsection{The spectral question}

The flat twisted representative of Sec.~\ref{sec:twist} has a strictly
positive free Faddeev--Popov gap, Eq.~\eqref{eq:FPgap}.  This should be
contrasted with the older observation that center-vortex configurations can
generate additional zero or near-zero Faddeev--Popov modes
\cite{Greensite:2004ur,Maas:2005ym}.  The contrast is useful: it shows that the
global flux and the vortex profile play distinct roles.

Let $A_\mu^{\mathrm v}$ be a center-vortex configuration in the same fixed
't~Hooft-flux sector.  We propose to study
\begin{equation}
 \operatorname{Spec}
 \mathcal{M}_B[A^{\mathrm v}]
 \label{eq:vortexspec}
\end{equation}
and compare it with
\begin{equation}
 \operatorname{Spec}
 \mathcal{M}_B[0].
 \label{eq:flatspec}
\end{equation}
The sharp question is whether the vortex profile closes the twist-induced gap
and drives one or more eigenvalues toward the boundary of $\Omega_B$.
This formulation avoids asking whether ``the vortex sector is on the horizon''
without specifying a configuration.

\subsection{\texorpdfstring{$\mathbb{R}^2\times T^2$}{R2 x T2} as a controlled bridge}

Yang--Mills theory on $\mathbb{R}^2\times T^2$ with a minimal 't~Hooft flux
provides a natural setting for this calculation.  In the small-$T^2$ regime,
semiclassical center-vortex configurations with fractional topological charge
can be treated in a controlled weak-coupling expansion
\cite{Tanizaki:2022ngt}.  Their characteristic action is
\begin{equation}
 S_{\mathrm v}
 =\frac{8\pi^2}{Ng^2},
 \label{eq:vortexaction}
\end{equation}
and their semiclassical weight takes the form
\begin{equation}
 \zeta_{\mathrm v}
 \sim K_{\mathrm v}
 \exp\!\left(-\frac{8\pi^2}{Ng^2}\right),
 \label{eq:vortexfugacity}
\end{equation}
where $K_{\mathrm v}$ denotes the fluctuation prefactor.

%A closed analytic expression for the required self-dual $\mathbb{R}^2\times T^2$ vortex profile is not presently available.  Its existence and continuum behavior were established through numerical constructions of twisted classical solutions \cite{GonzalezArroyo:1998ez,Montero:1999by,Montero:2000pb}; the modern semiclassical treatment likewise uses this numerically established saddle \cite{Tanizaki:2022ngt}.  Consequently the transition matrices determine the flat spectrum in Eq.~\eqref{eq:flatspec}, but do not determine Eq.~\eqref{eq:vortexspec}.  Completing the vortex part requires a numerical profile, a specified twisted Landau-gauge representative, and the corresponding Faddeev--Popov Hessian.

\revone{The original semiclassical treatment of self-dual center vortices
on $\mathbb{R}^2\times T^2$ was based on numerical constructions
of twisted classical solutions
\cite{GonzalezArroyo:1998ez,Montero:1999by,Montero:2000pb,
Tanizaki:2022ngt}.
More analytic information is available in an Abelianized regime.
Hayashi and Tanizaki related monopole semiclassics on
$\mathbb{R}^3\times S^1$ to the center-vortex description on
$\mathbb{R}^2\times T^2$ with 't~Hooft flux in center-stabilized
Yang--Mills theory \cite{Hayashi:2024yjc}.
On the covering space, the twist produces a chain of BPS and
Kaluza--Klein monopoles, the constituents of KvBLLY calorons,
by cyclically permuting the monopole species.
For $SU(2)$, this is an alternating BPS/KK chain.
In $\mathcal{N}=1$ super Yang--Mills theory, explicit holonomy
and dual-photon profiles were obtained in this regime, with
their relation expressing Abelian self-duality
\cite{Hayashi:2024psa}.
These results provide an analytic monopole-vortex description
in the Abelianized regime, rather than a closed-form fully
non-Abelian vortex profile for arbitrary torus aspect ratios.}

\revone{The transition matrices alone determine the flat spectrum in
Eq.~\eqref{eq:flatspec}, but do not specify the vortex background
entering Eq.~\eqref{eq:vortexspec}.
Evaluating the latter requires an explicit gauge-field profile
in a specified twisted Landau gauge, obtained numerically
or through a controlled analytic approximation.}

Write
\begin{equation}
 A_\mu=A_\mu^{\mathrm v}+a_\mu.
 \label{eq:vortexfluct}
\end{equation}
The next calculational target is the low-lying spectrum of
$\mathcal{M}_B[A^{\mathrm v}]$ as a function of the two compactification
lengths and of the vortex profile.  Three outcomes are logically possible.
The vortex may generate exact zero modes and lie on the sectorwise horizon;
it may generate parametrically small but nonzero modes and approach the
horizon in a scaling limit; or it may remain separated from the horizon even
though it dominates the semiclassical twisted partition function.  Each
outcome gives a distinct statement about the relation between the vortex and
Gribov mechanisms.

Only after this spectral problem is understood should the fluctuation
prefactor in Eq.~\eqref{eq:vortexfugacity} be reorganized with the
Gribov-restricted measure.  The decisive comparison is then between two
representations of the same gauge-invariant quantity:
\begin{equation}
 \left.
 \frac{Z[B]}{Z[0]}
 \right|_{\text{twisted-sector RGZ}}
 \quad\text{and}\quad
 \left.
 \frac{Z[B]}{Z[0]}
 \right|_{\text{vortex semiclassics}}.
 \label{eq:tworeps}
\end{equation}
Agreement would quantify the proposed bridge.  Disagreement would identify
which approximation fails and is equally informative.

\section{Practical route to a quantitative test}
\label{sec:numerics}

The Gaussian calculation is now reducible to three inexpensive numerical
objects.  The massive part is the Bessel sum in
Eq.~\eqref{eq:DeltaGammaTfinal}; for nonzero screening scales it converges
exponentially and can be truncated at a fixed winding length.  The massless
shape factor $\mathcal G_N$ is obtained from the primed scalar Laplacian on
the two ordinary tori appearing in Eq.~\eqref{eq:GNdef}, using either an
Epstein-zeta implementation or the heat-kernel integral
Eq.~\eqref{eq:zetaG}.  Finally, the constant-gluon correction depends only on
$K_0$.

\subsection{Numerical checks and a fixed-parameter \texorpdfstring{$SU(3)$}{SU(3)} example}
\label{subsec:numericalexample}

We performed three independent implementation checks before using the Bessel
sum numerically.  First, for $N=2$, $3$ the direct color-shifted heat kernel, the
enlarged-torus heat kernel, and the Poisson-resummed winding heat kernel agree
to better than $1.3\times10^{-13}$ for the sample range
$t/L^2=0.02$, $0.05$, $0.1$, $0.2$, $0.5$, $1$.  Second, a direct proper-time integration of
the enlarged-torus heat-kernel difference agrees with
Eq.~\eqref{eq:DNfinal} at relative accuracy $4.1\times10^{-10}$ for
$\mathfrak mL=1.3$ and at the $10^{-12}$ level for $\mathfrak mL=2$.  Third,
the regulated massless limit in Eq.~\eqref{eq:GNmasslesslimit}, evaluated directly
from the heat kernel and extrapolated in $(\mathfrak mL)^2$, gives
\begin{equation}
 \mathcal G_2\big|_{\mu_{\mathrm R}L=1}=4.69381,
 \qquad
 \mathcal G_3\big|_{\mu_{\mathrm R}L=1}=10.9577.
 \label{eq:numericalGN}
\end{equation}
Repeating the calculation at $L/L_0=0.7$ and $1.8$ reproduces the logarithmic
rescaling in Eq.~\eqref{eq:GNscaling} to better than $3\times10^{-14}$.
These checks separately test the fractional momentum shifts, the removed
sublattice, the winding projector, and the zero-mode subtraction.

For an illustrative physical scale, we use the central values of the
continuum-extrapolated $SU(3)$ tree-level RGZ fit of
Ref.~\cite{Dudal:2010tf},
\begin{equation}
 M^2=2.15~\mathrm{GeV}^2,
 \qquad
 m^2=-1.81~\mathrm{GeV}^2,
 \qquad
 \lambda^4=4.16~\mathrm{GeV}^4.
 \label{eq:numericalRGZparameters}
\end{equation}
In our convention the last quantity is $2g^2N\gamma^4$; hence
$M^2m^2+\lambda^4\simeq0.269~\mathrm{GeV}^4$.  The two roots are
\begin{equation}
 \mu_\pm^2
 =0.170\pm0.490\rmi~\mathrm{GeV}^2,
 \qquad
 \mu_\pm
 =0.587\pm0.417\rmi~\mathrm{GeV},
 \label{eq:numericalroots}
\end{equation}
where the square roots have positive real parts.  The imaginary part predicts
an asymptotic oscillation period of approximately $2.97~\mathrm{fm}$.
For the same parameters,
\begin{equation}
 K_0=0.125~\mathrm{GeV}^2>0,
 \qquad
 Q_{\mathrm{RGZ}}(z)
 =z^2+(0.340~\mathrm{GeV}^2)z+0.269~\mathrm{GeV}^4>0
 \label{eq:numericalpositivity}
\end{equation}
for $z\geq0$.  Thus the real Gaussian integral satisfies the convergence
conditions stated below Eq.~\eqref{eq:Kzero}, despite the complex
factorization masses.

\begin{figure}[t]
 \centering
 \includegraphics[width=0.78\linewidth]{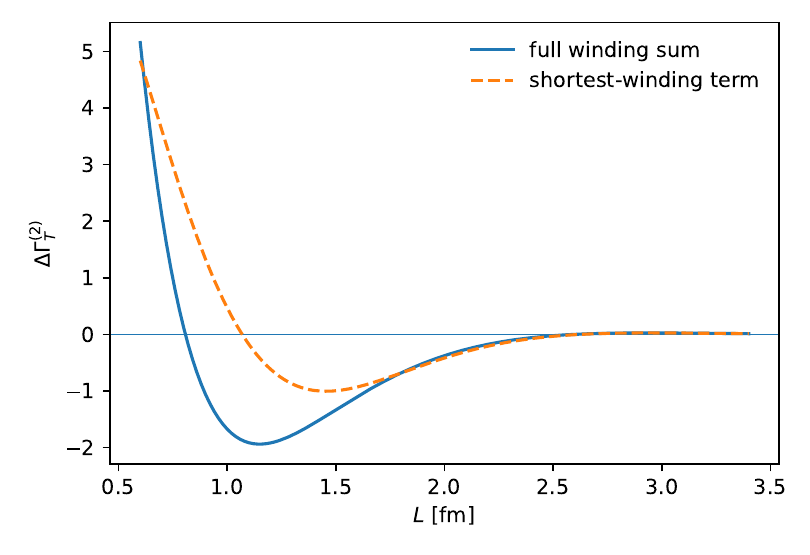}
 \caption{Fixed-parameter transverse response for an isotropic $T^4$ with
 $N=3$ and the RGZ parameters in Eq.~\eqref{eq:numericalRGZparameters}.  The
 solid curve is the complete winding sum in Eq.~\eqref{eq:DeltaGammaTfinal};
 the dashed curve retains only the shortest windings in
 Eq.~\eqref{eq:leadingisotropic}.}
 \label{fig:deltaGammaNumerical}
\end{figure}

Figure~\ref{fig:deltaGammaNumerical} evaluates the full winding sum on an
isotropic torus.  The response changes sign and then approaches zero through
a damped oscillation, while the shortest-winding approximation becomes
accurate once the box is sufficiently large.  This is a concrete realization
of the complex-pole behavior discussed in Sec.~\ref{subsec:largeL}; it is not
a self-consistent solution of the sectorwise RGZ gap equations.
The full sums in both figures were truncated to the four-dimensional sphere
$\bm\ell^2\leq18^2$.  Increasing the cutoff to $22^2$ over
$0.5~\mathrm{fm}\leq L\leq3.5~\mathrm{fm}$ changes
$\Delta\Gamma_T^{(2)}$ by at most $4.4\times10^{-10}$ and the complete
$\lambda^4$ source by at most $7.1\times10^{-9}~\mathrm{GeV}^{-4}$.

\begin{figure}[t]
 \centering
 \includegraphics[width=0.78\linewidth]{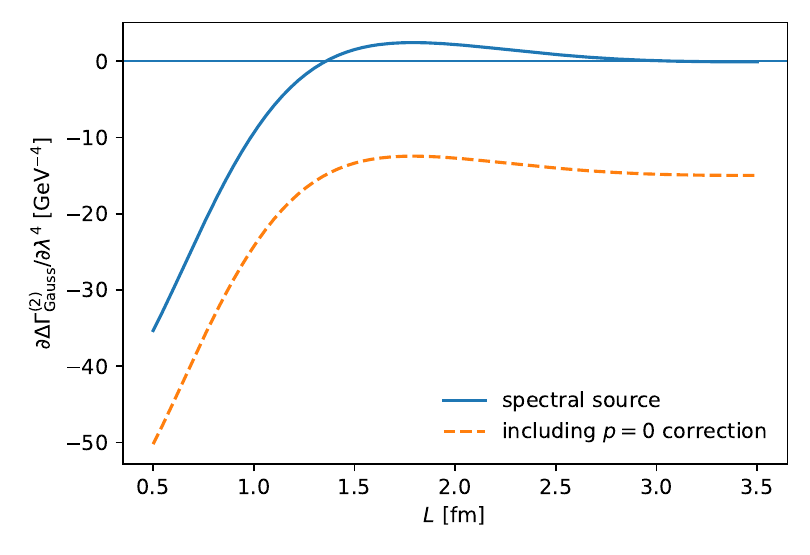}
 \caption{Finite-volume source for the horizon parameter at the fixed
 parameters of Eq.~\eqref{eq:numericalRGZparameters}.  The solid curve is the
 spectral Bessel contribution, while the dashed curve includes the untwisted
 constant-gluon correction.  The latter is order one in the effective action
 and therefore order $1/V_4$ in the effective-potential density.}
 \label{fig:horizonSourceNumerical}
\end{figure}

The same parameter choice also gives an immediate numerical test of the
horizon source.  Figure~\ref{fig:horizonSourceNumerical} separates the Bessel
spectral contribution in Eq.~\eqref{eq:dGammaTdLambda} from the full Gaussian
source in Eq.~\eqref{eq:dGammaFullLambda}.  The spectral term vanishes
exponentially, whereas the $p=0$ correction tends to the order-one constant
$-(N^2-1)/[2(M^2m^2+\lambda^4)]\simeq-14.9~\mathrm{GeV}^{-4}$.  As a
contribution to the effective-potential density this constant is suppressed
by $1/V_4$, as anticipated in Sec.~\ref{subsec:gapequations}.  As an
independent check of Eqs.~\eqref{eq:dGammaFullLambda},
\eqref{eq:dGammaFullm}, and \eqref{eq:dGammaFullM}, centered finite
differences of the complete parameter-dependent part of
Eq.~\eqref{eq:masterGaussian} at $L=1~\mathrm{fm}$ agree with the analytic
sources at relative accuracy $1.6\times10^{-8}$ or better.
The values at $L=1~\mathrm{fm}$ are collected in
Table~\ref{tab:numericalsources}.  The massless determinant and the stabilizer
do not enter these derivatives when their normalization is chosen
independently of the RGZ mass parameters.

\begin{table}[tb]
 \centering
 \caption{Fixed-parameter Gaussian response and stationary-equation sources
 at $L=1~\mathrm{fm}$ for Eq.~\eqref{eq:numericalRGZparameters}.}
 \label{tab:numericalsources}
 \begin{tabular}{lrr}
 \toprule
 Quantity & Numerical value & Mass dimension \\
 \midrule
 $\Delta\Gamma_T^{(2)}$ & $-1.67204$ & $0$ \\
 $\partial\Delta\Gamma_{\mathrm{Gauss}}^{(2)}/\partial M^2$
  & $54.2324$ & $-2$ \\
 $\partial\Delta\Gamma_{\mathrm{Gauss}}^{(2)}/\partial m^2$
  & $-43.8434$ & $-2$ \\
 $\partial\Delta\Gamma_{\mathrm{Gauss}}^{(2)}/\partial\lambda^4$
  & $-24.2751$ & $-4$ \\
 \bottomrule
 \end{tabular}
\end{table}

The massless regulator check has therefore also been completed at the level of
the primed Gaussian determinant.  Its absolute contribution to the physical
ratio remains inseparable from the common zero-mode/stabilizer normalization
identified in Eq.~\eqref{eq:stabilizermeasure}; the numerical value in
Eq.~\eqref{eq:numericalGN} should not be interpreted as an independently
normalized observable.
The ancillary Wolfram Language programs supplied with the source reproduce
the spectral identities, massless extrapolation, figures, tabulated data,
cutoff test, and finite-difference checks reported in this subsection.

The next step is the sectorwise RGZ vacuum.  For a chosen renormalized
$U_{\mathrm{RGZ}}$, one solves Eq.~\eqref{eq:gapeqgeneric} in the twisted and
untwisted sectors and compares
\begin{equation}
 x_i^{(B)}-x_i^{(0)}.
 \label{eq:paramdifference}
\end{equation}
The linear estimate Eq.~\eqref{eq:linearshift} can be checked directly.  For
the horizon equation the source term is already given explicitly by
Eq.~\eqref{eq:dGammaFullLambda}.  Varying $L_3$ and $L_4$ independently while
keeping $L_1$ and $L_2$ large isolates the response to the cycles carrying
the center twist.

To turn the Gaussian spectral answer into an absolutely normalized twisted
partition function, the residual global gauge symmetry should be fixed in a
common convention.  A particularly clean benchmark would be a lattice
calculation of the same flat-sector Gaussian ratio, where the Haar measure
fixes the zero-mode normalization unambiguously.  Matching that result to
$\Delta\Gamma_{\mathrm{stab}}$ would remove the only normalization constant
not determined by the continuum Hessian.

For the vortex-saddle stage, the relevant numerical observable is not only
the smallest eigenvalue.  One should track the low-mode density, the
localization of eigenfunctions, and their overlap with the source entering
the Zwanziger horizon functional as the vortex profile is varied.  A recent
configurationwise analysis emphasizes that a vanishing Faddeev--Popov
eigenvalue need not by itself make the source-sandwiched inverse singular
\cite{Tedesco:2026bsh}.  For a normalized low mode $\psi_k$, define the source
column and its visibility by
\begin{align}
 (m_{\mu d}[A])^a(x)
 &\equiv g f^{a\ell d}A_\mu^\ell(x),
 \nonumber\\
 \nu_k[A]
 &\equiv \sum_{\mu,d}
 \left|\braket{\psi_k\vert m_{\mu d}[A]}\right|^2.
 \label{eq:sourcevisibility}
\end{align}
For a degenerate critical multiplet, the basis-independent diagnostic is the
compressed visibility operator $P_cV_AP_c$, where $V_A=T_AT_A^\dagger$ and
$T_A$ has columns $m_{\mu d}[A]$.  At a simple isolated crossing, a nonzero
limiting value of $\nu_k$ fixes the residue of the singular horizon response.
Vanishing visibility only at the endpoint is not sufficient for boundedness;
the critical block must be checked along the approaching family.

The cleanest output would therefore be the joint spectral and visibility flow
\begin{equation}
 \left(
  \lambda_k[A^{\mathrm v}(\alpha)],
  \nu_k[A^{\mathrm v}(\alpha)]
 \right),
 \qquad
 0\le\alpha\le1,
 \label{eq:spectralflow}
\end{equation}
interpolating from the flat twisted representative at $\alpha=0$ to the
vortex saddle at $\alpha=1$.  A zero crossing or an accumulation of low modes
would show how a dynamical vortex approaches the sectorwise Gribov horizon,
while the visibility data determine whether that spectral approach is seen by
the Zwanziger horizon source.

\section{Discussion and conclusion}
\label{sec:conclusion}

We have organized the vortex--Gribov comparison around a gauge-invariant
twisted partition function and carried the flat-sector RGZ calculation to the
end of the Gaussian spectral problem.  The basic distinction remains crucial:
the background $\mathbb{Z}_N^{[1]}$ 2-form field labels a global sector, the
Gribov region constrains gauge fields in that sector, and a dynamical
center-vortex saddle is a particular configuration.  None of these three
objects should be identified by definition.

For a minimal irreducible twist on $T^4$, the adjoint spectrum is exactly
solvable.  The twist removes continuous global-color zero modes and produces
the free Faddeev--Popov gap in Eq.~\eqref{eq:FPgap}.  The shifted spectrum can
also be rewritten as the scalar spectrum of an enlarged torus with the
ordinary-torus sublattice removed, Eq.~\eqref{eq:twistedtraceTN}.  This makes
the global nature of the twist explicit and leads immediately to the
universal winding projector in Eq.~\eqref{eq:CNresult}.

The massive transverse RGZ determinant is the Bessel sum
Eq.~\eqref{eq:DeltaGammaTfinal}.  Its large-volume behavior is exponentially
small whenever the RGZ screening scales have positive real parts, and
complex-conjugate poles can produce damped oscillatory corrections.  This
part alone therefore forgets the center twist in the thermodynamic limit.

The main additional result of the present completion is
Eq.~\eqref{eq:masterGaussian}.  Once the full quadratic measure is integrated,
the Zwanziger localizing determinants cancel and no additional massive
oscillator determinant remains.  The longitudinal gluon and Faddeev--Popov
ghost instead leave the massless primed determinant $\mathcal G_N$, and the
absence of a twisted $p=0$ mode requires the explicit constant-gluon
correction proportional to $\ln K_0$.  The same calculation yields the
closed Bessel-sum sources for all three RGZ mass parameters in
Eqs.~\eqref{eq:dGammaFullLambda}--\eqref{eq:dGammaFullM}.

The one piece not fixed by the local Gaussian Hessian is the relative
normalization of global gauge stabilizers.  The untwisted flat connection is
reducible and has a continuous $SU(N)$ stabilizer, whereas the
irreducibly twisted connection has only the discrete center.  Priming the
Faddeev--Popov determinant removes the zero eigenvalues but does not specify
the corresponding orbit-volume measure.  This is precisely where a common
microscopic normalization is required.  In particular, the logarithmic
large-volume behavior of $\mathcal G_N$ should not be promoted to a physical
statement about $Z[B]/Z[0]$ before it is combined with that zero-mode measure.
The compensating renormalization-scale and volume dependences are fixed by
Eqs.~\eqref{eq:Gammaoscmuscaling} and \eqref{eq:GammaoscLscaling}, providing
direct matching conditions for any microscopic prescription.
The result is therefore stronger and sharper than the earlier transverse-only
calculation, but it is not a derivation of the fully normalized physical
partition-function ratio.

The remaining dynamical question is local rather than global.  On
$\mathbb{R}^2\times T^2$, one can place an actual semiclassical center-vortex
saddle in a fixed flux sector and follow the Faddeev--Popov spectrum from the
flat twisted representative to that saddle.  The twist-induced gap makes the
test sharp: if the vortex profile closes it, one obtains a direct spectral
bridge to the Gribov horizon; if it does not, the relation between the two
confinement pictures is necessarily more indirect.

The immediate quantitative tasks are therefore well separated.  The first is
to fix the global stabilizer normalization and solve a chosen renormalized
RGZ effective potential with the finite-volume source derived here.  The
second is the joint vortex-saddle spectral and source-visibility flow in
Eqs.~\eqref{eq:sourcevisibility} and \eqref{eq:spectralflow}.  Together they
provide two independent,
falsifiable probes of how center flux, the Gribov restriction, and dynamical
vortices are related.

\subsection{Observable-first perspective on the physical state space}

There is a broader question behind the use of gauge-invariant observables as primary probes of nonperturbative gauge dynamics.  In the perturbative formulation of gauge theory, the Kugo--Ojima/BRST construction~\cite{Kugo:1979gm} provides an exceptionally coherent organization of the physical state space: one first introduces a local and covariant gauge-fixed field theory on an enlarged indefinite-metric state space and then identifies physical states through BRST cohomology.  This field-first construction simultaneously preserves locality of the basic variables and provides a systematic cancellation of unphysical degrees of freedom.

It is less clear, however, to what extent this hierarchy of concepts should be regarded as fundamental beyond perturbation theory.  The global definition of gauge fixing is obstructed by Gribov copies, while lattice implementations of a global BRST symmetry encounter the Neuberger problem~\cite{Neuberger:1986vv,Neuberger:1986xz}.  These difficulties suggest that one may instead regard the gauge-invariant observable algebra as the primary nonperturbative object and reconstruct the physical state space from its representations.

Let $\mathcal A_{\mathrm{obs}}$ denote the algebra of gauge-invariant observables and let $\omega$ be a positive normalized state on $\mathcal A_{\mathrm{obs}}$.  The GNS construction~\cite{Segal:1947,Haag:1996} produces a representation
\begin{equation}
(\pi_\omega,\mathcal H_\omega,\Omega_\omega),
\end{equation}
such that
\begin{equation}
\omega(A)
=
\langle
\Omega_\omega,
\pi_\omega(A)\Omega_\omega
\rangle,
\qquad
A\in\mathcal A_{\mathrm{obs}}.
\end{equation}
If the von Neumann algebra
\begin{equation}
\mathcal M_\omega
=
\pi_\omega(\mathcal A_{\mathrm{obs}})''
\end{equation}
has a nontrivial center, its central decomposition further resolves the representation into factorial components,
\begin{equation}
\mathcal H_\omega
\simeq
\int_X^\oplus
\mathcal H_\xi\,\rmd\mu(\xi),
\qquad
\pi_\omega
\simeq
\int_X^\oplus
\pi_\xi\,\rmd\mu(\xi).
\end{equation}
This decomposition should be distinguished from, but is closely related to, the classification of superselection sectors as inequivalent representations of the quasilocal observable algebra satisfying appropriate localization criteria~\cite{Doplicher:1971,Doplicher:1974,Haag:1996}.

From this viewpoint, an important conceptual problem is whether the physical Hilbert space obtained from the observable algebra is equivalent, in a genuinely nonperturbative theory, to the space obtained from BRST cohomology in a gauge-fixed enlarged state space.  Perturbation theory strongly suggests such an equivalence in its domain of validity.  Nonperturbatively, however, the two constructions begin from different global structures, and their equivalence should not be assumed without further input.

Hamiltonian lattice gauge theory~\cite{Kogut:1974ag,Donnelly:2011hn,Casini:2013rba} provides a concrete arena in which this tension becomes operational.  After imposing Gauss law and restricting directly to the gauge-invariant Hilbert space, the physical degrees of freedom generally cease to possess a naive local tensor-product decomposition.  For a spatial bipartition one should not expect, in general,
\begin{equation}
\mathcal H_{\mathrm{phys}}
=
\mathcal H_R\otimes
\mathcal H_{\bar R}.
\end{equation}
Boundary electric fluxes, centers of regional observable algebras, and the associated superselection data replace this naive factorization.  Closely related phenomena appear in gauge-invariant encodings used in Hamiltonian formulations and quantum simulation: eliminating gauge redundancy can turn a manifestly local constrained theory into a theory of nonlocally encoded physical variables.

This observation raises a useful distinction.  The explicit nonlocality of a particular set of gauge-invariant variables may be an artifact of the chosen encoding~\cite{Zohar:2015jnb,Pardo:2023}.  The obstruction to naive Hilbert-space factorization and the existence of boundary or charge superselection sectors, by contrast, can be intrinsic properties of the observable theory.  Algebraic quantum field theory sharpens this distinction further: locality may remain exact at the level of observables even when charged fields or operators creating different sectors necessarily possess string-like or otherwise noncompact localization~\cite{Buchholz:1982}.

We therefore regard the comparison between the field-first and observable-first constructions as a structural question rather than merely a choice of gauge or computational formalism.  A possible nonperturbative hierarchy is
\begin{align}
&\text{observable algebra}
\\ &\qquad\longrightarrow
\text{states and GNS representations}
\\ &\qquad\longrightarrow
\text{sector decomposition}
\\ &\qquad\longrightarrow
\text{field realization},
\end{align}
instead of taking a gauge-fixed field algebra and its BRST cohomology as the primary starting point.  Such a hierarchy would sacrifice some of the manifest elegance and locality of the Kugo--Ojima construction at the level of field variables, but it may make global gauge structure, boundary sectors, and nonperturbative superselection data more transparent.

Whether these two routes reconstruct the same physical theory, and under which assumptions they do so, is in our view an important open problem connecting continuum gauge fixing, Hamiltonian lattice gauge theory, quantum simulation, and algebraic quantum field theory.  In the present context, this viewpoint motivates treating gauge-invariant sector-dependent quantities as primary observables and regarding a gauge-fixed nonperturbative formulation as a representation of their dynamics rather than as their definition.

\section*{Acknowledgements}
\revone{I would like to thank Prof.\ Hiroshi Suzuki, who first told me about the Gribov problem when I was an undergraduate, in response to a question that arose when I was first learning gauge theory. The question has remained with me ever since, and it has taken me more than ten years to formulate it in the form presented in this work.}

This work was partially supported by Japan Society for the Promotion of Science (JSPS)
Grant-in-Aid for Scientific Research Grant Number JP25K17402.
I acknowledge the RIKEN Special Postdoctoral Researcher Program
and RIKEN FY2025 Incentive Research Projects.

\appendix

\section{Poisson resummation of the shifted determinant}
\label{app:poisson}

For completeness, we derive Eq.~\eqref{eq:DNbeforecolor}.  For a shift vector
$\bm\delta$ define
\begin{align}
 \mathcal{D}_{\bm\delta}(\mathfrak m)
 ={}&\sum_{\bm n\in\mathbb{Z}^4}
 \ln\!\left[
 \sum_{\mu=1}^4
 \left(\frac{2\pi(n_\mu+\delta_\mu)}{L_\mu}\right)^2
 +\mathfrak m^2
 \right]
 \nonumber\\
 &-\sum_{\bm n\in\mathbb{Z}^4}
 \ln\!\left[
 \sum_{\mu=1}^4
 \left(\frac{2\pi n_\mu}{L_\mu}\right)^2
 +\mathfrak m^2
 \right].
 \label{eq:Ddelta}
\end{align}
The additive constant in the Schwinger representation cancels in the
difference.  Applying Poisson resummation in each direction,
\begin{align}
 &\sum_{\bm n\in\mathbb{Z}^4}
 \exp\!\left[
 -t\sum_\mu
 \left(\frac{2\pi(n_\mu+\delta_\mu)}{L_\mu}\right)^2
 \right]
 \nonumber\\
 &\hspace{8mm}
 =\frac{V_4}{(4\pi t)^2}
 \sum_{\bm\ell\in\mathbb{Z}^4}
 \exp\!\left(-\frac{\rho_{\bm\ell}^2}{4t}\right)
 \rme^{2\pi\rmi\bm\ell\cdot\bm\delta}.
 \label{eq:Poisson4d}
\end{align}
The $\bm\ell=0$ term cancels between the shifted and unshifted sectors.  The
remaining proper-time integral is
\begin{equation}
 \int_0^\infty\rmd t\,
 t^{-3}
 \exp\!\left(
 -\mathfrak m^2t-\frac{\rho^2}{4t}
 \right)
 =\frac{8\mathfrak m^2}{\rho^2}
 K_2(\mathfrak m\rho).
 \label{eq:Besselintegral}
\end{equation}
Consequently,
\begin{equation}
 \mathcal{D}_{\bm\delta}(\mathfrak m)
 =-\frac{V_4\mathfrak m^2}{2\pi^2}
 \sum_{\bm\ell\neq0}
 \left(\rme^{2\pi\rmi\bm\ell\cdot\bm\delta}-1\right)
 \frac{K_2(\mathfrak m\rho_{\bm\ell})}
 {\rho_{\bm\ell}^2}.
 \label{eq:Ddeltafinal}
\end{equation}
Summing Eq.~\eqref{eq:Ddeltafinal} over the adjoint shifts in
Eq.~\eqref{eq:SU_N_momenta} gives Eq.~\eqref{eq:DNbeforecolor}.

\section{Color character of the irreducible twist}
\label{app:color}

The color sum in Eq.~\eqref{eq:CNdef} follows from the elementary finite
Fourier transform
\begin{align}
 &\sum_{r,s=0}^{N-1}
 \exp\!\left[
 \frac{2\pi\rmi}{N}(\ell_3s-\ell_4r)
 \right]
 \nonumber\\
 &\hspace{12mm}
 =N^2
 \delta_{\ell_3=0\ (\mathrm{mod}\ N)}
 \delta_{\ell_4=0\ (\mathrm{mod}\ N)}.
 \label{eq:finiteFT}
\end{align}
Removing the singlet $(r,s)=(0,0)$ gives
\begin{equation}
 \sum_{(r,s)\neq(0,0)}
 \rme^{2\pi\rmi(\ell_3s-\ell_4r)/N}
 =N^2\delta_N(\ell_3)\delta_N(\ell_4)-1,
 \label{eq:adjchar}
\end{equation}
where $\delta_N(\ell)=1$ if $\ell\in N\mathbb{Z}$ and zero otherwise.
Subtracting one for each of the $N^2-1$ untwisted adjoint components gives
Eq.~\eqref{eq:CNresult}.

This result can be viewed as the adjoint character of the pair of twist
eaters evaluated on a winding $(\ell_3,\ell_4)$.  Its simplicity is the reason
the $SU(N)$ generalization of the finite-volume determinant is no
harder than the $SU(2)$ example.

\section{Gaussian determinant algebra}
\label{app:gaussian}

For completeness, we collect the exact Landau-gauge determinant counting
behind Eq.~\eqref{eq:Gammapermodefull}.  For a nonzero scalar eigenvalue
$z=p^2$, decompose
\begin{equation}
 A_\mu=A_\mu^T+\frac{p_\mu}{\sqrt z}\,a_L,
 \qquad
 p_\mu A_\mu^T=0.
 \label{eq:ATL}
\end{equation}
There are $d-1$ transverse real components and one longitudinal real
amplitude $a_L$.  The gauge-fixing delta function is
\begin{equation}
 \delta(p\mathbin{\cdot}A)
 =z^{-1/2}\delta(a_L),
 \label{eq:deltalandauappendix}
\end{equation}
while a complex Faddeev--Popov ghost pair gives
\begin{equation}
 \int\rmd c\,\rmd\bar c\,
 \rme^{-\bar c zc}
 \propto z.
 \label{eq:ghostdetappendix}
\end{equation}
The bosonic and Grassmann Zwanziger normalization determinants cancel with
identical power $d(N^2-1)$ as shown in
Eq.~\eqref{eq:auxmultiplicity}; their mixing with $A_\mu^T$ is already
contained in $K_T(z)$.  The nonzero-mode factor is therefore
\begin{equation}
 Z_p^{(2)}
 \propto
 z^{1/2}K_T(z)^{-(d-1)/2},
 \label{eq:Zlandauappendix}
\end{equation}
which is Eq.~\eqref{eq:Zpermodefull}.  As a check, when $K_T(z)=z$ the
effective action reduces to $(d-2)\ln z/2$, the determinant of the expected
$d-2$ massless physical polarizations.  The argument also shows why no
finite-$\alpha$ longitudinal continuation of the horizon kernel is required
for the Gaussian counting used in this paper.

\section{Massless shape determinant and scale dependence}
\label{app:massless}

Let $\zeta_T(s)$ denote the spectral zeta function of the scalar Laplacian on
a flat torus with its zero mode omitted.  The standard analytic continuation
obeys
\begin{equation}
 \zeta_T(0)=-1.
 \label{eq:zetazero}
\end{equation}
Under $L_\mu\to aL_\mu$, every nonzero eigenvalue scales as $a^{-2}$, so
\begin{equation}
 \zeta_{aT}(s)=a^{2s}\zeta_T(s).
 \label{eq:zetascaling}
\end{equation}
Since $\Ln\Det'\Delta=-\zeta_T'(0)$, it follows that
\begin{equation}
 \Ln\Det'\Delta_{aT}
 =\Ln\Det'\Delta_T+2\ln a.
 \label{eq:detScaling}
\end{equation}
Applying this once to $T_N^4$ and $N^2$ times to $T^4$ proves
Eq.~\eqref{eq:GNscaling}.

The appearance of a scale dependence in a primed determinant difference with
unequal zero-mode multiplicities is not paradoxical.  A determinant with
zero modes removed has inherited dimensions, and the missing dimensions are
supplied by the integration measure for the zero modes.  In the present
problem that measure is precisely part of
$\Delta\Gamma_{\mathrm{stab}}$.  This is why the combination entering a
fully normalized partition function cannot be specified by
$\mathcal G_N$ alone.

\section{Scope of the novelty claim and historical map}
\label{app:history}

The phrase ``link not yet made'' in Table~\ref{tab:historical-structure} is a
deliberately narrow bibliographic statement.  It does not assert that the
twisted adjoint momentum lattice, effective enlarged volume, Poisson
resummation, Gribov restriction, RGZ kernel, one-form-symmetry background, or
vortex spectral problem was individually unknown.  In particular, the
color-momentum construction and twisted-box perturbation theory are established
in Refs.~\cite{GarciaPerez:2014vt,Bribian:2019wqj}.  The link claimed here is
the combined construction: the gauge-invariant twisted ratio $Z[B]/Z[0]$ is
taken as the primary target; the Faddeev--Popov and RGZ measures are defined
sectorwise on the nontrivial bundle; the complete Gaussian determinant keeps
the massless primed determinant and the stabilizer mismatch;
%and the resulting flat-sector calculation is separated from, then compared with, a dynamical vortex saddle through spectral and source-visibility flow.
\revone{and a joint spectral and source-visibility flow is proposed
to compare the chosen flat representative with a dynamical
vortex saddle.}
We have not found
this combined construction or result in the cited twisted-bundle, RGZ, or
vortex-spectral literature.  The claim concerns this conjunction, not priority
for any one of its ingredients.

\begingroup
\footnotesize
\setlength{\tabcolsep}{2.3pt}
\renewcommand{\arraystretch}{1.08}
\begin{longtable}{
>{\raggedright\arraybackslash}p{0.105\linewidth}
>{\raggedright\arraybackslash}p{0.215\linewidth}
>{\raggedright\arraybackslash}p{0.285\linewidth}
>{\raggedright\arraybackslash}p{0.285\linewidth}}
\caption{Historical development of the ingredients entering the present
construction.  The last column identifies the link that was not the object of
the corresponding line of work; it should not be read as a claim that the
individual ingredients were previously unknown.}
\label{tab:historical-structure}\\
\toprule
Period & Line of development & What was established & Link not yet made \\
\midrule
\endfirsthead
\toprule
Period & Line of development & What was established & Link not yet made \\
\midrule
\endhead
\midrule
\multicolumn{4}{r}{\textit{Continued on the next page}} \\
\endfoot
\bottomrule
\endlastfoot
1978--1992
& Gribov geometry on compact configuration spaces
\cite{Gribov:1977wm,Singer:1978dk,Killingback:1984,vanBaal:1992}
& Global obstruction to gauge fixing, the Gribov horizon and fundamental
modular region, and explicit Gribov ambiguities on compact tori.  Twisted gauge
transformations also enter the topology of configuration space.
& No formulation in terms of a one-form-symmetry background $B$ and no
sectorwise RGZ calculation of the gauge-invariant ratio $Z[B]/Z[0]$. \\

1979--1982
& 't~Hooft flux and twisted bundles
\cite{tHooft:1979rtg,vanBaal:1982}
& Gauge-invariant electric/magnetic flux sectors on a hypertorus and the
description of nontrivial bundles by twisted boundary conditions.
& The twisted sector was not combined with a Gribov restriction or an RGZ
spectral determinant. \\

2005--2006
& Center vortices and the Faddeev--Popov spectrum
\cite{Greensite:2004ur,Maas:2005ym}
& Center-vortex configurations were related to enhanced low-lying
Faddeev--Popov modes and to special loci of the Gribov horizon.
& The analysis was configuration-first: a vortex background was inserted and
its gauge-fixed spectrum was studied, rather than starting from a
gauge-invariant flux-sector partition function. \\

2008--2023
& RGZ and background-field GZ/RGZ
\cite{Dudal:2008sp,Kroff:2018,Dudal:2023rki}
& The refined Gribov--Zwanziger framework introduced infrared screening
scales, while later background-field formulations made it possible to study
background-dependent effective potentials and deconfinement.
& These constructions were not organized around a nontrivial 't~Hooft bundle
and its symmetry-twisted partition-function ratio. \\

2014--2019
& Twisted-box perturbation theory and volume reduction
\cite{GarciaPerez:2014vt,Bribian:2019wqj}
& The twist-eater color basis, fractional color momentum, effective enlarged
periods, infrared cutoff, and perturbative finite-volume sums were developed
for irreducibly twisted tori.
& These tools were not combined with a sectorwise Gribov restriction and the
complete RGZ Gaussian response of $Z[B]/Z[0]$. \\

2015--2026
& Generalized symmetry and twisted partition functions
\cite{Gaiotto:2014kfa,Hayashi:2026ht}
& The center symmetry was formulated as a $\mathbb Z_N^{[1]}$ one-form symmetry
with background two-form field $B$, and the torus ratio $Z[B]/Z[0]$ was proposed
as a gauge-invariant criterion for center-vortex condensation.
& A Gribov/RGZ realization of this gauge-invariant observable, including the
twisted Faddeev--Popov spectrum and finite-volume determinant, was not supplied.
\\

%2022--2026
%& Semiclassical center vortices on compactified geometries \cite{Tanizaki:2022ngt,Hayashi:2026ht}
%& Controlled center-vortex saddles on compactified Yang--Mills geometries provide a configuration-level semiclassical description of center flux.
%& Their Faddeev--Popov spectral and source-visibility flow relative to the flat representative of the same fixed 't~Hooft-flux sector remains a separate dynamical question. \\
\revone{2022--2026}
&
\revone{Semiclassical center vortices on compactified geometries
\cite{Tanizaki:2022ngt,Hayashi:2024yjc,
Hayashi:2024psa,Hayashi:2026ht}}
&
\revone{Controlled center-vortex saddles provide a semiclassical
description of center flux.
Analytic monopole--vortex connections have also been
established in Abelianized regimes.}
&
\revone{Their Faddeev--Popov spectral flow relative to the flat
representative of the same fixed 't~Hooft-flux sector
remains a separate dynamical question.}
\\

Present work
& Gauge-invariant target and sectorwise RGZ realization
& We take $Z[B]/Z[0]$ as the primary observable, solve the adjoint spectrum in a
fixed irreducible 't~Hooft-flux sector, derive its exact enlarged-torus
representation and finite-volume RGZ response, and separate the global twist
from an actual dynamical vortex configuration.
& The remaining problems are the common zero-mode/stabilizer normalization,
self-consistent sectorwise RGZ vacuum selection, and the Faddeev--Popov
spectral and source-visibility flow around an actual vortex saddle. \\
\end{longtable}
\endgroup

\bibliographystyle{utphys}
\bibliography{ref}

\end{document}